\documentclass[structabstract]{aa}  
\pdfoutput=1
\usepackage{graphicx}
\usepackage{longtable,lscape}
\usepackage{txfonts}

\def\mstar{$M_{\rm *}$}

\def\arcsec{\hbox{$^{\prime\prime}$}}
\def\degr{\hbox{$^\circ$}}
\def\micron{$\mu$m}
\def\kms{km s$^{-1}$}

\def\cmc{cm$^{-3}$}

\def\ergscm{ergs s$^{-1}$ cm$^{-2}$}

\def\msun{M$_{\odot}$}
\def\msunpyr{M$_{\odot}$ yr$^{-1}$}
\def\zsun{Z$_{\odot}$}

\def\mstar{M$_*$}
\def\iab{$I_{\rm AB}$}
\def\iab{$I_{\rm AB}$}
\def\d4000{$D_{\rm 4000}$}
\def\halpha{\ifmmode {\rm H{\alpha}} \else $\rm H{\alpha}$\fi}
\def\hbeta{\ifmmode {\rm H{\beta}} \else $\rm H{\beta}$\fi}
\def\hgamma{\ifmmode {\rm H{\gamma}} \else $\rm H{\gamma}$\fi}
\def\hda{\ifmmode {\rm H{\delta}_{\rm A}} \else $\rm H{\delta}_{\rm A}$\fi}
\def\hdelta{\ifmmode {\rm H{\delta}} \else $\rm H{\delta}$\fi}
\def\oii{[O\,{\sc ii}]$\lambda$3727}

\def\oiiib{[O\,{\sc iii}]$\lambda$5007}

\def\niib{[N\,{\sc ii}]$\lambda$6584}
\def\nii{[N\,{\sc ii}]$\lambda\lambda$6548,6584}
\def\sii{[S\,{\sc ii}]$\lambda\lambda$6717,6731}
\def\siir{[S\,{\sc ii}]$\lambda$6717/$\lambda$6731}
\def\neiii{[Ne\,{\sc iii}]$\lambda$3869}

\def\civ{C\,{\sc iv}$\lambda$1550}

\def\oiiisoii{[O\,{\sc iii}]/[O\,{\sc ii}]}
\def\siissii{[S\,{\sc ii}]6717/[S\,{\sc ii}]6731}

\def\nn2{$N2$}
\def\rr23{$R_{\rm 23}$}
\def\oo32{$O_{\rm 32}$}
\def\oo2ne3{$O_{\rm 2Ne3}$}
\begin{document}
   \title{MASSIV: Mass Assemby Survey with SINFONI in VVDS\thanks{This work is based mainly 
    on observations collected at the European Southern Observatory
    (ESO) Very Large Telescope, Paranal, Chile, as part of the Programs 
    179.A-0823, 177.A-0837, 78.A-0177, 75.A-0318, and 70.A-9007. This work also benefits from data products 
    produced at TERAPIX and the Canadian Astronomy Data Centre as part of the 
    Canada-France-HawaiiTelescope Legacy Survey, a collaborative project of NRC 
    and CNRS.}}
   \subtitle{I. Survey description and global properties of the $0.9 < z < 1.8$ galaxy sample}

\titlerunning{MASSIV. I. Survey description and global properties of the galaxy sample}

\author{
T. Contini \inst{1,2}
\and B. Garilli \inst{3}
\and O. Le F\`evre \inst{4}
\and M. Kissler-Patig \inst{5}
\and P. Amram \inst{4}
\and B. Epinat \inst{1,2}
\and J. Moultaka \inst{1,2}
\and L. Paioro \inst{3}
\and J. Queyrel \inst{1,2}
\and L. Tasca \inst{4}
\and L. Tresse \inst{4}
\and D. Vergani \inst{6}
\and C. L\'opez-Sanjuan \inst{4}
\and E. Perez-Montero \inst{7}
         }

\offprints{T. Contini, \email{contini@ast.obs-mip.fr}}

   \institute{
  Institut de Recherche en Astrophysique et Plan\'etologie (IRAP), CNRS, 14, avenue Edouard Belin, F-31400 Toulouse, France \and
  IRAP, Universit\'e de Toulouse, UPS-OMP, Toulouse, France \and
  IASF-INAF, Via Bassini 15, I-20133, Milano, Italy \and
  Laboratoire d'Astrophysique de Marseille, Universit\'e de Provence,
  CNRS, 38 rue Fr\'ed\'eric Joliot-Curie, F-13388 Marseille Cedex 13,
  France \and
  ESO, Karl-Schwarzschild-Str.2, D-85748 Garching b. M\"unchen,
  Germany \and
  INAF-Osservatorio Astronomico di Bologna, via Ranzani 1, I-40127,
  Bologna, Italy \and
  Instituto de Astrof\'\i sica de Andaluc\'\i a - CSIC Apdo. 3004
  E-18080, Granada, Spain
               }

   \date{Received ; accepted }

  \abstract
   {}
   {Understanding how galaxies evolve and assemble their mass across cosmic time is still a fundamental unsolved issue. To get insight into the various processes of galaxy mass assembly, the Mass Assembly Survey with SINFONI in VVDS (MASSIV) aims at probing the kinematical and chemical properties of a significant and representative sample of high-redshift ($0.9\leq z \leq 1.8$) star-forming galaxies.}
{This paper presents the selection function, the observing strategy and the global properties of the MASSIV sample. This sample contains 84 star-forming galaxies, selected from the VIMOS VLT Deep Survey (VVDS) and observed with the SINFONI integral-field spectrograph at the VLT. We present the redshift distribution, and derive the stellar masses and SED-based star formation rates (SFR). Integrated metallicities and the presence of type-2 AGNs are investigated using composite 1D spectra built from VIMOS and SINFONI observations.}
   {The MASSIV selection function, based on star formation criteria (\oii\ emission-line strength up to $z\sim 1.5$ and colors/UV absorption lines at higher redshifts), provides a good representation of ``normal" star-forming galaxies with SED-based SFRs between $5$ and $400$ \msunpyr\ in the stellar mass regime $10^9 - 10^{11}$ \msun.  
Analysis of typical emission-line ratios performed on composite spectra reveals that the contamination by type-2 AGNs is very low and 
that the integrated metallicity of the galaxies follows the well-known mass-metallicity relation.}
{The MASSIV sample has been built upon a simple selection function, fully representative of the star-forming galaxy population at $0.9 < z < 1.8$ for SFR $\geq5$ \msunpyr. Together with the size of the sample, the spatially-resolved SINFONI data therefore enables us to discuss global, volume averaged, galaxy kinematic and chemical properties all accross the mass and SFR range of the survey to derive robust conclusions on galaxy mass assembly over cosmological timescales. All the data published in this paper are publicly available at the time of publication following this link: {\tt http://cosmosdb.lambrate.inaf.it/VVDS-SINFONI}.}

   \keywords{Galaxies: evolution - Galaxies: high-redshift - Galaxies: kinematics and dynamics - Galaxies: starburst
               }

   \maketitle
%

\section{Introduction}

The details of evolution processes driving galaxy assembly are still largely unconstrained. 
During the last decade, major spectroscopic and multi-wavelength photometric surveys 
(eg. VVDS, COSMOS, DEEP2, GOODS, CFHTLS, etc) explored in depth the 
high-redshift universe, allowing to follow the evolution of large numbers of galaxies over cosmological timescales.
Even if these data enabled significant progress concerning 
the knowledge of the {\it global} properties (spectral energy distribution, colours, stellar mass and age, star 
formation rate, nebular metallicity, size, etc) of different galaxy populations in different environments, more {\it detailed} 
constraints are needed to understand the main physical processes involved in the formation and evolution of galaxies. 

Based on observational and/or theoritical  arguments, we know that galaxy merging (eg.\ Bournaud et al. 2007;  
Lin et al. 2008; Conselice et al. 2008, 2009; de Ravel et al. 2009; L\'opez-Sanjuan et al. 2011) is at play in the 
growth of galaxies. Recently, cold gas accretion along cosmic filaments (Kere{\v s} et al. 2005; 
Ocvirk et al. 2008; Genel et al. 2008, 2010; Dekel et al. 2009) has been emphasized as an efficient process to sustain high
star formation rates and explain the clumpy nature of high-$z$ disks. However observational signatures for cold accretion 
is still debated (e.g. Steidel et al. 2010; Le Tiran et al. 2011) and we do not know yet at which cosmic epoch 
one of these two processes (mergers and cold accretion) is dominant.

Spatially-resolved observations of high-redshift galaxies started to give some clues to answer
this question as they allow to probe both the internal properties (kinematics, distribution of metals, ISM physics, etc) 
and close environment of distant galaxies. Powerful integral field spectrographs mounted of the largest ground-based 
telescopes (eg.\ GIRAFFE and SINFONI on the VLT, OSIRIS on Keck) recently made such studies possible, giving access 
to the bright  rest-frame optical emission lines such as \halpha, \hbeta, \niib, \oiiib, and \oii\ in $z\sim 0.4-4$ star-forming galaxies.  
However, most of the previous surveys concentrated their efforts on the $z\geq2$ universe (SINS: F{\"o}rster Schreiber et al. 2006, 2009 -- OSIRIS: 
Law et al. 2007, 2009; Wright et al. 2007, 2009 -- LSD/AMAZE: Maiolino et al. 2008; Mannucci et al. 2009) with the exception 
of IMAGES who probed the kinematics of intermediate-$z$ galaxies (eg.\  Puech et al. 2008, 2010).  
However, much less is know in the $z\sim1-2$ redshift range, except some very recent studies on small ($\leq 30$ galaxies) samples (Wisnioski et al.\  2011; Mancini et al.\  2011), where we know that $L^*$ galaxies, which are the major contributors to the
cosmic star formation rate density at this epoch (Cucciatti et al. 2011), are experiencing major transformations.
Do cold gas accretion be still an efficient process to assemble galaxies or do major mergers be predominant as
it seems to be the case at later epochs (Yang et al. 2008, Kere{\v s} et al. 2009)?

MASSIV (Mass Assembly Survey with SINFONI in VVDS) tackles this issue by surveying a representative
sample of star-forming galaxies in the redshift range $z\sim1-2$. 
With the detailed information provided by SINFONI on individual galaxies, 
the key science goals of the MASSIV survey are to investigate in detail: 
(1) the nature of the dynamical support (rotation vs. dispersion) of high-$z$ 
galaxies, (2) the respective role of mergers (minor and/or major) and gas 
accretion in galaxy mass assembly, (3) and the process of gas exchange 
(inflows/outflows) with the intergalactic medium through the derivation of 
metallicity gradients.

The MASSIV sample includes 84 star-forming galaxies drawn from 
the VIMOS VLT Deep Survey (VVDS) in the redshift range $0.9 < z < 2.2$. 
We stress the importance of working on representative and statistically-significant samples. 
Indeed, the acquisition of high-$z$ galaxy kinematics is a long observational process which requires to target 
galaxies that are representative of the main physical processes occurring at $1 < z < 2$, in order to draw 
conclusions which are unbiased toward any specific population. 
We present the full MASSIV sample, focusing on the selection 
criteria, the observing strategy, and the global properties derived mainly 
from SED fitting and integrated VIMOS and SINFONI composite spectra. 
The analysis of a first sample of 50 MASSIV galaxies is presented 
in related papers focusing on the kinematical classification (Epinat et al. 2011), 
the evolution of scaling relations such as the baryonic Tully-Fischer relation (Vergani et al. 2011), and 
the spatially-resolved metallicity (Queyrel et al. 2011). 
The analysis of a sub-sample of 9 MASSIV galaxies observed during pilot runs in 2005-2006 has already been 
published in Epinat et al. (2009) and Queyrel et al. (2009). The analysis of the full MASSIV sample, including the 
other 34 galaxies which have been observed with SINFONI, will be published in subsequent papers after 
the completion of data reduction and kinematical modeling.

The paper is organized as follows. 
The selection of MASSIV targets is described in Section~\ref{sample}. The observing strategy 
with SINFONI is presented in Section~\ref{obstrat} and includes a discussion on the advantages and drawbacks 
of using adaptive optics with the higher spatial sampling compared to seeing-limited observations. 
In Section~\ref{global}, we present the global properties (redshift, stellar mass, and 
star formation rate) of the MASSIV sample mainly derived from the SED fitting 
and we discuss how well it represents the $z\sim 1-2$ star-forming galaxy population. 
The integrated VIMOS and SINFONI composite spectra are shown and discussed in 
Section~\ref{spectra}. The MASSIV sample global properties are 
compared to those of other Integral Field Unit (IFU) spectroscopic samples at similar 
redshifts in Section~\ref{comparison}. The paper is summarized
in Section~\ref{conclusion}. 

Throughout this paper, we assume a standard $\Lambda$-CDM cosmology, i.e. $h=0.7$,
$\Omega_{\mathrm{m}}=0.3$ and $\Omega_{\Lambda}=0.7$ (Spergel et al. 2003). 
For this cosmology, 1\arcsec corresponds to $\sim 8$ kpc at $z\sim 1-2$. Magnitudes
are given in the AB-based photometric system, unless explicitly stated otherwise.

\section{The MASSIV sample}
\label{sample}

	\subsection{Parent VVDS catalogues and target selection criteria}
	\label{selcrit}

We have used the VIMOS VLT Deep Survey (VVDS) sample to select galaxies with known spectroscopic redshifts accross the peak of cosmic star formation activity ($z\sim 1-2$). 

The VVDS is composed of three $I$-band selected surveys totalizing about
47\,000 spectra of galaxies, quasars, and stars from $z\sim0$ to
$z\simeq5$: (1) a Wide survey ($17.5 \le I_{\rm AB}\le
22.5$; Garilli et al. 2008), (2) a Deep survey ($17.5 \le I_{\rm AB}\le
24.0$; Le F\`evre et al. 2004, 2005), and (3) an Ultra-Deep survey
($23.00 \le i'_{\rm AB}\le24.75$; Le F\`evre et al., in prep.).  Spectra
have a spectroscopic identification at a confidence level higher than
$\sim$ 50\%, 60\%, 81\%, 97\% and 99\%, corresponding to the VVDS
quality flags 1, 9, 2, 3 and 4.

The VVDS is a complete magnitude-selected sample avoiding the biases linked to \textit{a priori} color selection techniques. This sample offers the advantage of combining a robust selection function and secure spectroscopic redshifts. The latter are necessary to engage into long integration times on single galaxies with SINFONI being sure to observe the H$\alpha$ line away from bright OH night-sky emission lines. From the existing VVDS dataset, we have access today to a unique sample of more than 4400 galaxies in the redshift domain $0.9 < z < 2$ with accurate and secure spectroscopic redshifts (Le F\`evre et al. 2005). 
This sample being purely $I$-band limited, it contains both star-forming and passive galaxies distributed over a wide range of stellar masses, enabling us to easily define volume-limited sub-samples. 

For the MASSIV survey, we have defined a sample of 84 VVDS star-forming galaxies at $0.9 < z < 2.2$ suitable for SINFONI observations. Three selection criteria have been applied successively. 
First, the MASSIV targets were selected to be star-forming galaxies, based on their \oii\ or UV flux. The selection was performed on the measured intensity of \oii\ emission line in the VIMOS spectrum (see Lamareille et al. 2009; Vergani et al. 2008) or, for the cases where the \oii\ emission line was out of the VIMOS spectral range (i.e. for $z\gtrsim 1.5$), on the UV flux based on their observed photometric $UBVRIK$ spectral energy distribution and/or UV rest-frame spectrum. 
The star formation criteria ensure that the brightest rest-frame optical emission lines (mainly H$\alpha$ and \niib, or in a few cases \oiiib) used to probe kinematics and chemical abundances, will be observed with SINFONI in the NIR $J$ and $H$ bands. Among these star-forming galaxy candidates, we have further restricted the sample taking 
into account two important observational constraints: the observed wavelength of H$\alpha$ line has to fall at least 9\AA\ away from strong OH night-sky lines\footnote{this limit allows the detection of \halpha\ far enough from the galaxy dynamical center assuming rotating disks with a maximum circular velocity lower than 200 \kms}, in order to avoid heavy contamination of the galaxy spectrum by sky subtraction residuals; and the \halpha\ line had to lie in the $J$ or $H$ band of SINFONI, thus prohibiting the redshift range $z=1.049-1.192$. Finally, 90\% of MASSIV galaxies have been selected to be observable at higher spatial resolution with the adaptive optics (AO) system of SINFONI assisted with the Laser Guide Star (LGS) facility. 
In these cases, a bright star ($R < 18$ mag) close enough to the target ($d < 60\arcsec$) is needed for the 
zero-order tip-tilt corrections, as indicated in the SINFONI User Manual.  

In the following sub-sections, we detail the criteria used for the selection of MASSIV targets depending on the depth (Wide, Deep and Ultra-Deep) of the parent VVDS catalogue.

		\subsubsection{The Wide VVDS sample}

The VVDS Wide used VIMOS at the ESO VLT to target four separate fields widely distributed on the 
sky (RA= 02h, 10h, 14h, and 22h) and covering a total of 16 deg$^2$. The spectroscopic sample has been derived from an $I$-band selected photometric catalogue applying a pure apparent magnitude limit at \iab\ $= 22.5$. The final spectroscopic sample contains $\sim$20\,000 galaxies (broad-line QSOs are excluded on the basis of their peculiar spectral features) with a median redshift $z\sim 0.58$ (Garilli et al. 2008).  
The average target sampling rates of VVDS Wide are
between 20\% to 25\% over $17.5\leq I_{\rm AB}22.5$, and are dependent of
the magnitude (Garilli et al. 2008). The redshift success rates are
$\sim95,70,85,80$\% for the 02h, 10h, 14h, 22h, respectively.

MASSIV targets have been drawn from the VVDS-22h and VVDS-14h Wide fields. These fields contain a total of $\sim 1500$ galaxies  with a secure spectroscopic redshift over the redshift range $0.9 < z < 1.5$. 

\begin{figure}
   \centering
   \includegraphics[bb=18 144 592 718,width=9cm]{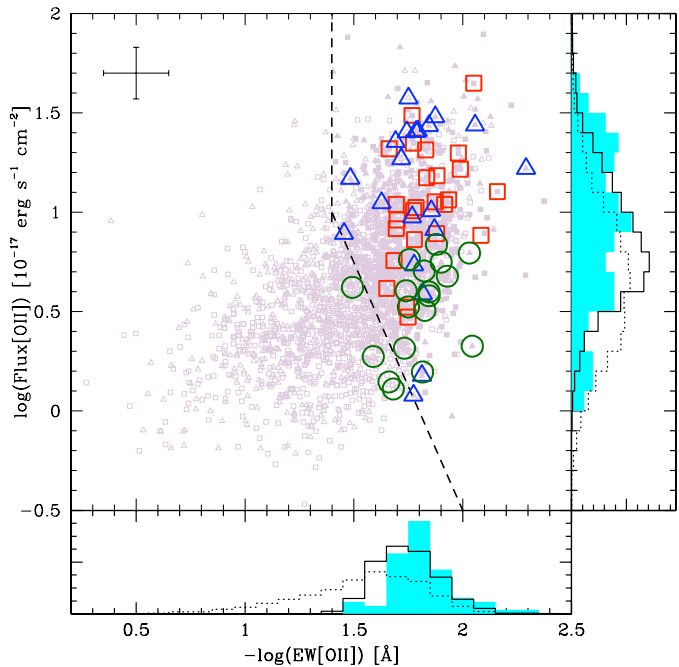}
      \caption{Selection of MASSIV $0.9 \lesssim z \lesssim 1.5$ star-forming galaxies with secure redshift 
and measured \oii\ emission line in the VVDS Wide (VVDS-14h and -22h, \iab\ $\leq 22.5$, small triangles) 
and Deep/Ultra-Deep VVDS-02h (\iab\ $\leq 24/24.75$, small squares) fields for SINFONI follow-up 
observations. Among these galaxies we selected those showing strong enough \oii\ emission line for \halpha\ to be easily detected in the NIR, and satisfying the OH and AO/LGS criteria (filled symbols, see text for details). 
The 63 galaxies selected for the MASSIV survey in the redshift range $0.9 \lesssim z \lesssim 1.5$ are indicated as blue triangles (VVDS-14h and -22h Wide fields), red squares (VVDS-02h Deep field), and green circles 
(VVDS-02h Ultra-Deep field). The distribution of \oii\ equivalent width (EW) and flux are 
projected onto the x- and y-axis respectively. These histograms are normalized to the total number of objects. 
Empty-dot histograms are for the full VVDS Wide, Deep, and Ultra-Deep samples with measured \oii\ emission line ($N=2397$, median $EW= -38\AA$, median flux $= 4.4\times 10^{-17}$\ergscm). 
Empty-solid histograms are for the VVDS Wide, Deep, and Ultra-Deep sample galaxies complying with the selection criteria (\oii\ strength, OH sky lines and redshift range, see text for details, $N=728$, median $EW= -54\AA$, median flux $= 6.9\times 10^{-17}$\ergscm). 
Filled cyan histograms are for the selected MASSIV galaxies ($N=63$, median $EW= -61\AA$, median flux $= 8.3\times 10^{-17}$\ergscm). 
The dashed line indicates the selection box of MASSIV targets based on \oii\ EW, flux and spectrum SNR (see text for details) The cross in the upper-left corner shows the typical error bars on \oii\ flux and EW.
              }
         \label{selection}
   \end{figure}

In order to select secure star-forming (SF) galaxies, we based our selection on the strength of  \oii\ emission lines measured in VIMOS spectra up to $z\sim 1.5$ (see Fig.~\ref{selection}), when this line leaves the VIMOS spectral domain. The choice of the lower limit on \oii\ equivalent width (EW) depends on the signal-to-noise ratio: it is equal to $-25\AA$ for S/N $\geq 10$ (this is the case for 42 galaxies or $\sim 66$\% of the full \oii-selected MASSIV sample) and 
$-40\AA$ for $6 \leq$ S/N $< 10$ (this is the case for 22 galaxies). Galaxies with a S/N $< 6$ on \oii\ EW were not considered. This limit on \oii\ EW ensures that the \halpha\ emission line will be detected with 
SINFONI with a sufficient S/N ratio in a reasonable exposure time. A visual check has also been performed on VIMOS spectra to verify that the \oii\ line was not affected by a sky subtraction residual. We further applied the OH sky lines and SINFONI bands criteria (see above). 

This selection strategy, based on \oii\ equivalent width instead of using simply a flux limit, has been built on past experiences, especially 
on the detection rate obtained after the two pilot runs with SINFONI used to define the best selection strategy for the MASSIV Large Program. 
For the pilot runs, we used a selection based both on \oii\ flux {\it and} equivalent width (see Fig. 1 in Epinat et al. 2009). After a careful 
inspection of both detected and undetected targets with SINFONI, we concluded that a selection based on \oii\ EW only allows to probe 
more "normal" star-forming galaxies by reaching SFR as low as a few \msunpyr\ without  decreasing significantly the detection rate.

A total of 224 galaxies finally satisfied the above mentioned  criteria in the VVDS-22h and -14h fields. 
In this sample, 183 galaxies have a bright star close enough and could in principle be observed with AO-LGS. 

Finally, 21 galaxies have been selected in the VVDS Wide sample for SINFONI follow-up observations. The 13 MASSIV targets in the VVDS-22h field 
have been selected randomly among these 183 galaxies, but 4 only were observed with the AO-LGS facility. The 8 additionnal targets selected  
in the VVDS-14h field do not have any bright star close enough to be observed with AO-LGS. 

		\subsubsection{The Deep VVDS sample}
		
The VVDS Deep survey used VIMOS at the ESO VLT to target data in the
VVDS-0226-04 (Le F\`evre et al. 2005) and VVDS-CDFS fields
(Le F\`evre et al. 2004).  In this study, we use the VVDS-0226-04 field only,
which covers 2200 arcmin$^2$ of sky area. It includes 9842 spectra
described in Le F\`evre et al. (2005), plus 2826 spectra acquired later
with the same set-up . Observations of several $z\ge1.4$
targets, with assigned VVDS quality flags of 0, 1, or 2, have been reapeated in order to
assess their real redshift distributions as detailled in Le~F\`evre et
al. (in prep).  The measurement of the target sampling and redshift
success rates are estimated in Ilbert et al. (2005) and Cucciati et al. (2011), the
later work is accounting for the repeated observations. In summary,
the average target sampling rate is $\sim24-25$\%, and is calculated 
as a function of the projection of the angular size of each target in the 
$x$-axis of the image.  The redshift success rate is estimated as a function of $I_{\rm AB}$ and
redshift; for instance at $z<1.5$, it is $\sim100$\% at $I_{\rm AB} \le
22.5$\%, and it smoothly decreases down to $\sim80$\% at the faintest
magnitudes.  Stellar masses have been derived for all galaxies with a
measured redshift (see eg. Walcher et al. 2008).  The
VVDS flux limit translates at different redshifts into different lower
luminosity limits. This, in turn, translates into a broad mass
selection cut at each redshift, reflecting the scatter in the
Mass$-$Luminosity relation. Meneux et al. (2008) have studied the stellar
mass incompleteness as a function of redshift.  At $z<1.5$, it is on
average 84\% complete in stellar mass for \mstar$>10^{9.5}$\msun.

MASSIV targets have been drawn from this Deep VVDS field which contains a total of $\sim 2600$ galaxies with secure spectroscopic redshift over the redshift range $0.9 < z < 2.0$. 

The criteria applied for the selection of secure star-forming galaxies is identical to the one used for the Wide VVDS catalogue, i.e. based on the strength of the \oii\ emission line. We further applied the OH sky lines and SINFONI bands criteria as for the Wide fields. 
A total of 490 galaxies satisfy these criteria in the VVDS-02h Deep field.  
In this sample, 300 galaxies have a bright star close enough and could in principle be observed with AO-LGS. 

Finally, 29 MASSIV targets have been selected randomly in this parent sample of 300 galaxies. Among these targets one is observed with the AO-LGS facility.

		\subsubsection{The Ultra-Deep VVDS sample}
\label{vvdsud}

The VVDS Ultra-Deep has assembled a sample of 1200 new targets over 576
arcmin$^2$ within the 2200 arcmin$^2$ sampled by the Deep survey. 
Thanks to its large spectral range ($\sim 3600-9350 \AA$) and long exposure time 
($\sim15$ hours), this unique spectroscopic survey enables one to securely identify 
high-redshift galaxies up to $z\sim 5$, filling the so-called ``redshift desert" 
between $z=1$ and $2$. The
average target sampling rate of the Ultra-Deep survey is constant, i.e. $\sim4$\% at $23\le
I_{\rm AB}\le 24.75$. Combining the VVDS Deep and Ultra-Deep surveys
($17.5\le i'_{\rm AB} \le 24.75$), Cucciati et al. (2011) have derived the
following rates. The target sampling rate starts from $\sim20$\% at
$I_{\rm AB}\sim22.5$, reaches $\sim30$\% at $I_{\rm AB}\le23$, and then decreases to
2\% at $I_{\rm AB}\le24.75$. The redshift success rate is measured as a
function of magnitude and redshift. From $z=1.5$ to $z=2$, it
decreases from $\sim80$\% to $\sim65$\% (see Le F\`evre et al., in
prep.).

This Ultra-Deep VVDS catalogue has been used to select MASSIV galaxies in the redshift domain $z\sim 1.24-1.8$, in order to target \halpha\ in the $H$ band. 
It complements nicely the $\sim 1.2-1.5$ redshift range, already populated with sources drawn from the VVDS Wide and Deep, and allows us to extend the MASSIV 
sample up to $z\sim 1.8$ with a robust selection. 

The selection procedure to identify secure star-forming galaxies for the SINFONI follow-up is different from the one described above for the Wide and Deep VVDS catalogues. Indeed, when the MASSIV target selection has been performed, no systematic measurement of spectral features (emission and absorption lines) was applied to the Ultra-Deep sample. We thus selected galaxies with either a prominent \oii\ emission line (as measured with IRAF \textit{splot} task) or, for redshifts greater than $\sim 1.5$, a rest-frame UV spectrum typical of star-forming galaxies (i.e. with 
strong \civ\ absorption line and/or blue continuum, see Fig.~\ref{vimoscomposite}). We further applied the OH sky lines criterium as for the Wide and Deep fields restricting the VVDS Ultra-Deep sample to 86 galaxies fulfilling our criteria. In this sample, 55 galaxies have a bright star close enough and could be in principle observed with AO-LGS. 

Finally, 34 MASSIV targets (including 18 \oii-selected at $z < 1.5$) have been drawn from this parent sample of 55 galaxies. Among these targets 6 are observed with the AO-LGS facility.

		\subsubsection{The final MASSIV sample}

The full MASSIV sample is extracted from the three VVDS samples described above, for which the completeness functions are well-controlled. This sample, selected either on the \oii\ EW strength or on the UV intensity, is essentially limited in flux translating into a lower limit for the star formation rate (see Fig.~\ref{sfrsedvsmass}).  Figure~\ref{selection} illustrates the selection procedure based on \oii\ emission line strength for the Wide, Deep, and Ultra-Deep VVDS samples. The distribution of the 63 MASSIV targets, selected among the strongest \oii\ emitters (lower limit on $EW$ only) span a wide range of \oii\ flux (from $\sim 1.2$ to $45\times 10^{-17}$ \ergscm) and $EW$ (from $\sim -200$ to $-30\AA$). The median values of these distributions ($EW= -61\AA$, flux $= 8.3\times 10^{-17}$\ergscm) are slightly higher than the median values of the combined VVDS Wide, Deep and Ultra-Deep parent samples 
($EW= -54\AA$, flux $= 6.9\times 10^{-17}$\ergscm) 
complying with the selection criteria listed above (\oii\ strength and redshift range). It is important to stress that the expected star formation rates (ranging from $\sim8$ to $\sim1000$ \msunpyr\ with a median value of $41$ \msunpyr, and using the [OII]-based calibration of Argence \& Lamareille 2009) corresponding to these limits imposed on the observed \oii\ line strength are in very good agreement, taking the dust attenuation into account, with the SED-based star formation rates derived in Section~\ref{sedfit}.

The properties of the MASSIV sample are further described in the next sections focusing on the selection function (sect.~\ref{seleffects}) and the comparison with other major IFU surveys of high-$z$ galaxies (sect.~\ref{comparison}).

\section{SINFONI observations}
\label{obstrat}
	
	\subsection{Strategy}

The observations of the MASSIV sample were performed with SINFONI 
(Eisenhauer et al. 2003; Bonnet et al. 2004) mounted at the Cassegrain focus 
of the VLT UT4 telescope. 
Most of the data (90\%) have been collected in service mode during several observing periods between 
April 2007 and January 2011, as part of the ESO Large Program 179-A.0823. However, eleven galaxies of the 
MASSIV sample (including three with duplicated observations) were observed during two pilot runs with SINFONI 
(September 2005 and November 2006, in visitor mode). The analysis of nine of these galaxies has been 
published in Epinat et al. (2009) and Queyrel et al. (2009).    

To map the \halpha, \niib, and \sii\ emission lines, or \oiiib\ and \hbeta\ for 
four galaxies of the MASSIV sample, we used the $J$ or $H$ gratings, depending on the redshift of the sources, 
which have a nominal FWHM spectral resolutions of $R\sim 1900$ and $2900$ respectively\footnote{The nominal spectral resolution is slightly lower with the 0.05\arcsec/pixel scale used for some AO-assisted observations (see p.\ 68 of the SINFONI User Manual, version P89).}. 
The vast majority of the observations (74 galaxies) were carried out in seeing-limited mode with the
largest pixel scale of 0.125\arcsec/pixel giving a Field-of-View (FoV) of 
8\arcsec$\times$8\arcsec. We observed a total of eleven targets with adaptive optics (AO) assisted 
with the Laser Guide Star (LGS). 
For seven of them, we selected the intermediate
0.05\arcsec/pixel scale with FoV of 3.2\arcsec$\times$3.2\arcsec to
take full advantage of the gain in angular resolution provided
by the AO, achieving a FWHM resolution of $\sim 0.25$\arcsec. 
For the four other targets with AO data, we used the larger pixel scale 
as trade-off between enhanced angular resolution and sensitivity 
(see below for a detailled discussion).
The observing conditions were generally good, with clear to photometric sky
transparency and typical seeing at NIR wavelengths with
FWHM = $0.5-0.8$\arcsec. 

Depending on the FoV, we adopted two different observing
strategies to maximize the efficiency of observations. 
When the largest 8\arcsec$\times$8\arcsec\ FoV was used, i.e.\ for the 
vast majority of observations, we applied 
an "object nodding" technique by moving the target from one corner to 
the other of the FoV.   For the seven galaxies observed with AO-LGS at 
0.05\arcsec/pixel, we adopted the strategy of taking the sky frames away 
from the target because the source size did not allow nodding 
within the small 3.2\arcsec$\times$3.2\arcsec FoV.  In such cases, the telescope 
was pointing between the object (ÒOÓ) and adjacent sky regions (ÒSÓ) 
in an ÒO-O-S-O-O-O-O-S-O-OÓ pattern. 

The individual exposure times vary between 300s and 600s 
depending mainly on the grating $J$ or $H$ used for observations (individual 
exposures of 900s have been used for a few galaxies during the pilot runs, see 
Epinat et al. 2009). The total on-source integration times
range on average from 80 to 120 minutes. The total integration
times were driven by the \oii\ emission line intensity as measured 
in VIMOS spectra or, when this information was not available, 
by the $I$-band magnitude of the target. 

In order to estimate the spatial PSF, we obtained exposures of a bright star close 
to the target (or tip-tilt stars for AO-LGS), used also to center the galaxy into the FoV 
with a blind offset. 
For flux calibration and atmospheric transmission correction, we observed 
telluric standard stars as part of the standard ESO calibration plan. 
For details on the data reduction procedures, we refer to Epinat et al. (2011).

\subsection{Consistency between AO and seeing-limited observations: VVDS220596913 as a test case}
\label{ao}

\begin{figure*}[t]
   \centering
   \includegraphics[width=18cm]{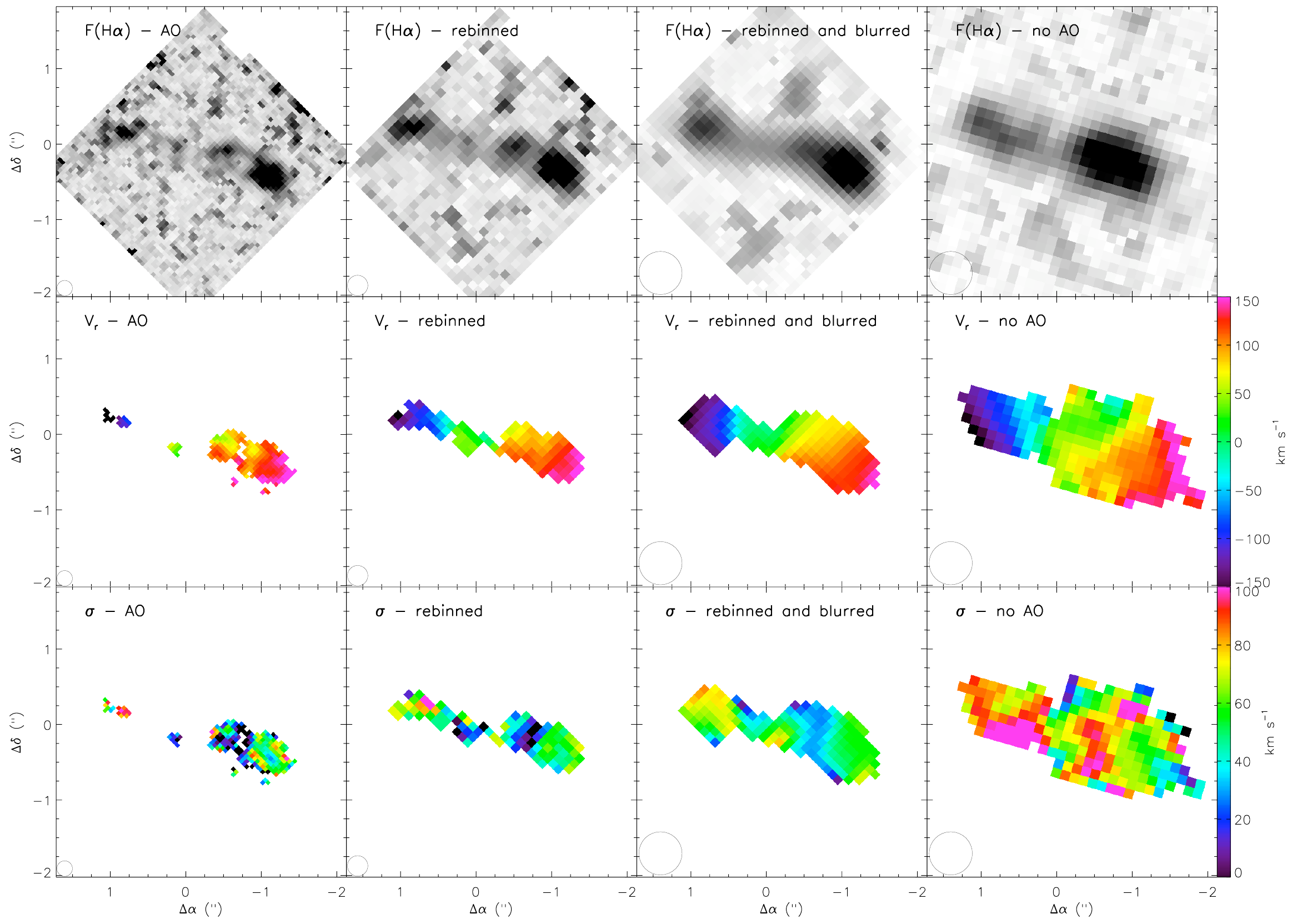}
\caption{\halpha\ maps and velocity fields of VVDS220596913 ($z\sim 1.27$) obtained with and without adaptive optics. From top to bottom: \halpha\ flux map, velocity field and velocity dispersion map  (corrected for spectral resolution). From left to right: AO-assisted SINFONI observation (pixel size = 0.05\arcsec), simulated observation with AO and a pixel size of 0.10\arcsec, simulated seeing-limited observation with a pixel size of 0.10\arcsec, seeing-limited SINFONI observation (pixel size = 0.125\arcsec).  The circle in the lower left corner of each image represents the effective FWHM spatial resolution (a two pixel Gaussian smoothing has been applied to each cube before map extraction). The color scale of the \halpha\  flux maps is arbitrary. A threshold of S/N per pixel $\geq 3$ is used for the display of velocity maps. North is up and East is left. }
\label{ao_vs_noao}
\end{figure*}

\begin{table*}[t]
\caption{VVDS220596913: physical parameters derived from SINFONI observations (with and without AO) and simulations.  Parameter $\sigma$ corresponds to the intrinsic velocity dispersion corrected for both instrumental broadening (spectral resolution) and beam smearing (see Epinat et al. 2011 for details)}
\label{table:1}
\centering                          
   \begin{tabular}{lcccc}
\hline\hline
Parameter  & SINFONI & Model & Model & SINFONI \\
      & AO & Binned & Binned and blurred & Seeing-limited\\
\hline
Spatial sampling [\arcsec] & 0.05 & 0.10 & 0.10 & 0.125 \\
Spatial resolution [\arcsec] & 0.183 & 0.183 & 0.573 & 0.573 \\
Exposure time [minutes] & 120 & $\cdots$ & $\cdots$ & 160 \\
\halpha\ flux [$\times 10^{-16}$ \ergscm] & 3.0 & $\cdots$ & $\cdots$ & 4.0 \\
\halpha\ size [\arcsec] & 2.6 & 2.5 & 2.7 & 3.4 \\
Position angle [\degr] & $247 \pm 1$ & $250 \pm 1$ & $248 \pm 1$ & $246 \pm 1$ \\
V$_{max}$ [\kms] & $141 \pm 2$ & $127 \pm 2$ & $155 \pm 2$ & $165 \pm 3$ \\
$\sigma$ [\kms] & $38 \pm 28$ & $41 \pm 24$ & $37 \pm 22$ & $66 \pm 22$ \\
\hline
   \end{tabular}
\end{table*}

As stated previously, even if the majority of MASSIV galaxies have been observed with SINFONI in natural seeing conditions, eleven galaxies of the sample benefit from AO-LGS observations in order to achieve a better spatial resolution. However, these AO observations were not performed always with the same spatial sampling. Indeed, the observing strategy we adopted with the small pixel scale ($0.05$\arcsec) at the beginning of the survey requires the acquisition of sky frames in separate exposures (see above). Thus, for a given total observing time, only 80\% can be spent on the object. To optimize the survey efficiency, we decided to modify our strategy for AO-assisted observations from Period 82, shifting to a larger pixel scale (from $0.05$\arcsec\ to $0.125$\arcsec). Thus, at the end, seven galaxies out of eleven have been observed with the smallest pixel scale. For an homogeneous analysis and interpretation of the resolved properties of MASSIV galaxies, it is 
then important to check whether the differences in the final spatial resolution/sampling achieved with the SINFONI datacubes have an impact on the derived galaxy 
properties (\halpha\ flux, velocity fields, etc). 

The galaxy VVDS220596913 at $z\sim 1.27$ is a perfect ``test case" for this purpose as it has been observed twice with SINFONI. A natural seeing observation has been first acquired in 2007 as part of a pilot run (see Epinat et al. 2009). This object was then re-observed in the frame of the MASSIV survey,  but this time using AO-LGS. Data from the pilot run have been re-reduced using the same method than for the MASSIV data (Epinat et al. 2011). In Table~\ref{table:1}, we summarize the observing conditions and derived parameters for the seeing-limited and AO-assisted SINFONI observations of VVDS220596913. Taking into account both the variation in pixel surface and the difference in exposure time between the two observations, we expect the signal-to-noise per pixel of the AO observation to be lower than in the seeing-limited one by a factor of 2.9. Using the \halpha\ flux maps shown in Figure~\ref{ao_vs_noao}, we measure a factor of $\sim 2.5$. This slight difference may be due to the fact that the signal is less diluted in AO observations and/or observations were not performed under the same photometric conditions.

From our AO observations with high spatial sampling ($0.05$\arcsec/pixel), it was possible to resample the data in order to infer what would be the gain in using AO with a larger pixel scale  with respect to non-AO observations and with respect to small-pixel AO observations. In order to avoid any interpolation, we used a spatial resampling of $0.10$\arcsec\ instead of $0.125$\arcsec. We also mimic seeing-limited observations by adding a blurring of the datacube to reach the same final spatial resolution of our non-AO observation but with a $0.10$\arcsec/pixel spatial sampling. We created datacubes from which we derived \halpha\ flux and kinematics maps after a two pixels FWHM spatial Gaussian smoothing, as is the case for the observed data. The limitation of these simulations is that the lowest surface brightness regions could be not detected at all, thus, rebinning the data would not be strictly equivalent to real observations. This is further supported by the fact that real and simulated non-AO observations are not strictly identical in the relative flux of the various components (see Figure \ref{ao_vs_noao}).

The gain of AO-assisted observations with the large pixel scale with respect to non-AO observations is clearly visible. Indeed, in the non-AO observation it is difficult to distinguish the four blobs/clumps that are clearly separated in AO-assisted observations independently of their pixel scale (see Figure \ref{ao_vs_noao}). Due to surface brightness detection limit, these clumps might be even more clearly detected with the large pixel scale (both fainter surface brightness limit and longer exposure time). However, the interpretation would still remain the same since we do not detect significant change in the velocity field which remains monotonic. 

We further compared the size of the \halpha\ emission, the flux in this line as well as, assuming a rotating disk model (see Epinat et al. 2011), the kinematics position angle of the major axis, the maximum rotation velocity and the mean velocity dispersion (see Table~\ref{table:1}). We find that all these quantities are lower in the AO observation. However, except for the velocity dispersion, they remain compatible within 25\%. This can be understood by the fact that the AO observation is less deep and thus misses some low-surface brightness regions. Using a similar signal-to-noise threshold, we end up with final maps that are slightly less extended than in the seeing-limited observations. Note that we also find a very good 
agreement between the various kinematics position angles of the major axis obtained from kinematics modeling.

We have perfomed advanced tests to better understand the larger variations observed in the velocity dispersion. First, it has to be noticed that the difference before correction for spectral PSF is only of the order of $10$ \kms. The accuracy of the wavelength calibration could thus be a cause for the broadening of lines in non-AO observations due to the nodding strategy. However, we have checked that the wavelength calibration cannot account for the observed difference (it could only account for $\sim 1$ \kms). The determination of the spectral PSF is also unlikely to explain alone the difference. One other explanation could be a spatial offset between various individual exposures since several components at various velocities can be summed up if spatial offsets are not correctly controlled. However, we would expect the same offsets in AO and non-AO observations. The accuracy of the offsets cannot be checked since no bright component is present in the data. Two other effects could explain the observed discrepancy in velocity dispersions: the lower SNR in the AO data and the presence of a sky line residual close to the \halpha\ line. The SNR has been probed using a single OB for the non-AO data, but is seems that the velocity dispersion remains at the same value, even if locally it can change by about $5$ to $10$ \kms. However it could be that low surface brightness regions are not observed in the AO observation. The final difference is probably a combination of all these effects. It has however to be mentionned that the final velocity dispersion remains compatible between the different configurations within the computed uncertainties.

Based on this test case, we conclude that the final resolution/sampling of SINFONI observations does not impact significantly the derived \halpha\ flux and velocity maps.    

\section{Global properties of the MASSIV sample}
\label{global}

	\subsection{Redshift distribution}
	
\begin{figure}[t]
   \centering
   \includegraphics[bb=18 144 592 718,width=9cm]{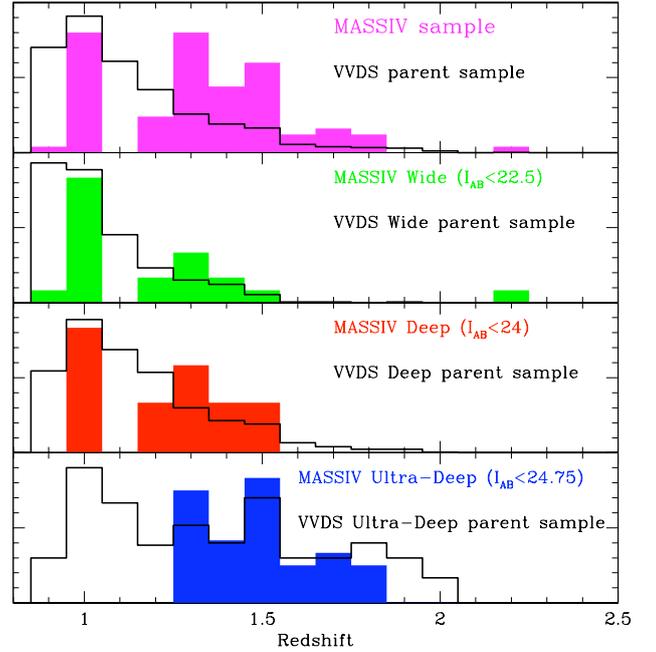}
      \caption{Redshift distribution of the MASSIV samples (filled histograms) compared with the parent (i.e. restricted to the $0.9-2.0$ redshift range) VVDS samples (white histograms). From top to bottom: complete (magenta), Wide (green), Deep (red) and Ultra-Deep (blue) MASSIV and VVDS samples. Histograms are arbitrarily normalized. The redshift distributions of MASSIV reflect the selection criteria, but are also importantly affected by the observability of the target emission lines (mainly \halpha, \oiiib\ in a few cases) in the NIR atmospheric bands (explaining the gap around $z\sim 1.1$) and between the OH night sky lines. MASSIV galaxies are distributed in the redshift range between $z\sim 0.94$ and $1.80$, with the exception of VVDS220148046 which has a redshift $z=2.24$ (see text for details).
                          }
         \label{zhist_comp}
   \end{figure}

The redshift distributions of the MASSIV sample are compared with those of the parent VVDS catalogues in Figure~\ref{zhist_comp}. Median values and associated dispersion are listed in Table~\ref{table:2}. MASSIV galaxies are distributed in the redshift range between $z\sim 0.94$ and $1.80$, with the exception of VVDS220148046 which has a redshift $z=2.24$. This galaxy was supposed to be in the same redshift range as other MASSIV targets but it appeared that the redshift derived from the VIMOS spectrum was wrong. The redshift of this galaxy is now secured with the measurement of \hbeta\ and \oiiib\ emission lines in the SINFONI spectra. The median redshift for the MASSIV sample is $z=1.33$. The lowest redshift bin ($z\sim 1$) is populated with galaxies selected in the Wide and Deep VVDS samples, whereas most of the galaxies at higher redshifts ($z > 1.5$) are selected from the Ultra-Deep VVDS sample only. The median redshift values for the MASSIV galaxies drawn from the Wide, Deep and Ultra-Deep VVDS samples are $z=1.05$, $z=1.28$, and $z=1.48$ respectively.
	
	\subsection{Physical parameters derived from SED fitting}
	\label{sedfit}
	
   \begin{figure}[t]
   \centering
   \includegraphics[bb=18 144 592 718,width=9cm]{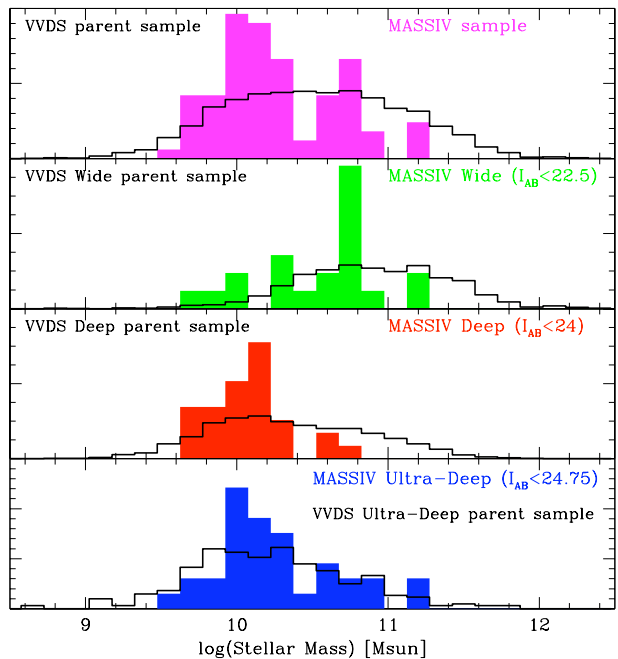}
      \caption{Stellar mass distribution of the MASSIV samples (filled histograms) compared with the parent VVDS samples (white histograms). From top to bottom: complete (magenta), Wide (green), Deep (red) and Ultra-Deep (blue) MASSIV and VVDS samples. Histograms are normalized to the total number of objects.
                          }
         \label{masshist}
   \end{figure}
	
   \begin{figure}[t]
   \centering
   \includegraphics[bb=18 144 592 718,width=9cm]{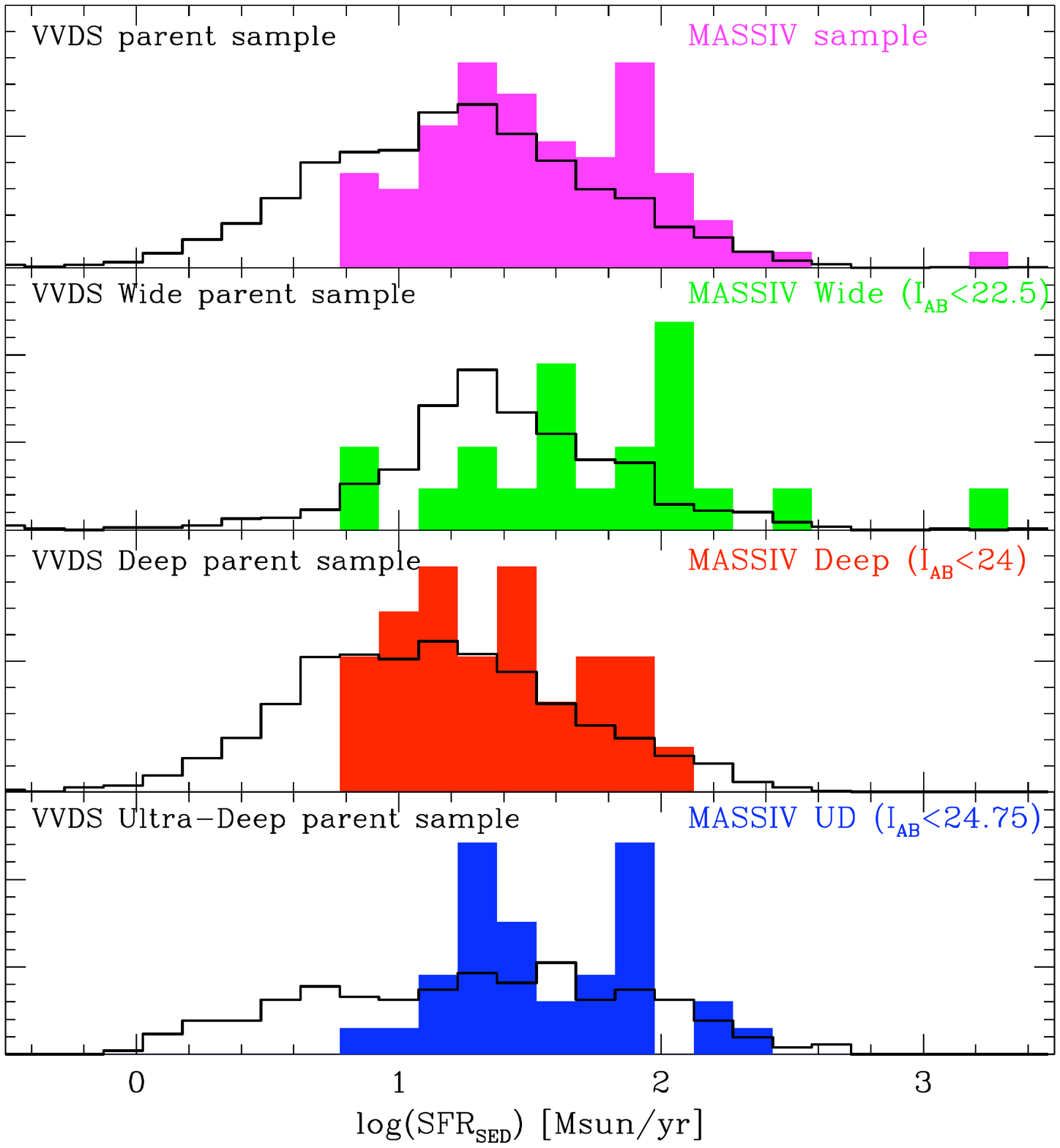}
      \caption{SED-derived SFR distribution of the MASSIV samples (filled histograms) compared with the parent VVDS samples (white histograms). From top to bottom: complete (magenta), Wide (green), Deep (red) and Ultra-Deep (blue) MASSIV and VVDS samples. Histograms are normalized to the total number of objects.
                          }
         \label{sfrsedhist}
   \end{figure}

   \begin{table*}
   \tabcolsep1.40mm
\caption{Statistics on the redshift and global physical parameters of the MASSIV and parent VVDS samples}             
\label{table:2}      
\centering                          
\begin{tabular}{l c r c r c r c r c r c r }        
\hline\hline                 
Sample &  \multicolumn{4}{c}{Redshift} & \multicolumn{4}{c}{log(Stellar Mass) [\msun]} & \multicolumn{4}{c}{log(SFR$_{\rm SED}$) [\msunpyr]}  \\    
 & \multicolumn{2}{c}{MASSIV} & \multicolumn{2}{c}{VVDS} & \multicolumn{2}{c}{MASSIV} & \multicolumn{2}{c}{VVDS} & \multicolumn{2}{c}{MASSIV} & \multicolumn{2}{c}{VVDS}  \\  
  & $Median\pm \sigma$ & $N$ &  $Median\pm \sigma$ & $N$ &  $Median\pm \sigma$ & $N$ &  $Median\pm \sigma$ & $N$ &  $Median\pm \sigma$ & $N$ &  $Median\pm \sigma$ & $N$ \\
\hline                        
Total & $1.33\pm0.13$ & 84 & $1.06\pm 0.13$ & 4446 & $10.15\pm 0.30$ & 84 & $10.51\pm 0.43$ & 4248 & $1.50\pm 0.31$ & 84 & $1.22\pm 0.35$ & 3023  \\
Wide & $1.05\pm 0.14$ & 21 & $1.00\pm 0.09$ & 1483 & $10.68\pm 0.24$ & 21 & $10.89\pm 0.34$ & 1409 & $1.80\pm 0.25$ & 21 & $1.36\pm 0.27$ & 783  \\
Deep & $1.28\pm 0.15$ & 29 & $1.09\pm 0.13$ & 2643 & $10.08\pm 0.14$ & 29 & $10.30\pm 0.38$ & 2537 & $1.37\pm 0.28$ & 29 & $1.12\pm 0.36$ & 1982  \\
Ultra-Deep & $1.48\pm 0.11$ & 34 & $1.33\pm 0.25$ & 320 & $10.15\pm 0.29$ & 34 & $10.21\pm 0.33$ & 302 & $1.50\pm 0.27$ & 34 & $1.33\pm 0.47$ & 258  \\
\hline                                   
\end{tabular}
\end{table*}
	
In addition to optical VIMOS spectroscopy, all VVDS fields have a
multi-wavelength photometric coverage: $B,V,R,I$ from the VIRMOS Deep
Imaging Survey (VDIS, Le F\`evre et al. 2004),  {\it u*, g', r', i', z'} from
the Canada-France-Hawaii Telescope Legacy Survey (CFHTLS, but for VVDS-22h and -02h only), 
$J$, $H$, and $K$ from UKIDSS, and a partial coverage in $J$ and $K$ bands from VDIS (Iovino et al. 2005). 
The VVDS Deep field has also been covered in the IRAC 3.6, 4.5, 5.8, and 8.0 \micron\ bands by Spitzer within the 
SWIRE survey (Lonsdale et al. 2003).

Making use of the optical spectra and of all the photometry available, 
we have fitted the spectral energy distribution (SED) of each
galaxy to obtain estimates of the stellar mass and star formation rate (SFR) 
of our sample and of the parent VVDS sample. 
The fitting of the SED has been performed using  the GOSSIP spectral energy
distribution modeling software (Franzetti et al.
2008):  photometric and optical spectroscopic data were fitted with a
grid of stellar population models, generated using the BC03
population synthesis code (Bruzual \& Charlot 2003), assuming
a set of ``delayed'' star formation histories (see Gavazzi
et al. 2002 for details), a Salpeter (1955) IMF with lower
and upper mass cutoffs of respectively 0.1 and 100 \msun, a
metallicity ranging from 0.02 and 2.5 solar metallicity and
a Large Magellanic Cloud reddening law (Pei 1992) with
an extinction $E(B-V)$ ranging from 0 to 0.3. The parameters
for the best-fitting model for each galaxy are taken as
the best fitting values for both the galaxy stellar mass and
SFR. On top of the best fitting values, GOSSIP computes
also the Probability Distribution Function (PDF), following
the method described in Walcher et al. (2008). The median
of the PDF and its confidence regions are then used to derive
a robust estimate of the parameter and associated uncertainty that
is to be determined. 

We have used all the photometric information available for each galaxy 
within the MASSIV sample. Each fit has been
visually inspected, and in few cases photometric points strongly
deviating from the global fit have been removed.
Within VVDS, spectroscopic data are flux calibrated at the 10--20\%
level (Le F\`evre et al. 2005). In order to fit them together with
photometry, we have renormalized spectroscopic data to the photometric 
$i'$-band magnitude. As MASSIV galaxies have been selected on the basis
of their star-forming activity, we have been able to further constrain
the fitting by imposing that the resulting model should not have a
$D_{4000}$ break larger than $1.25$ with some tolerances to account for 
the uncertainties on the measurement of $D_{4000}$ break. This is a reasonable value that excludes 
early-type {\em passive} galaxies but keeps the possibility to deal with an early-type 
{\em star-forming} galaxy (see e.g. Vergani et al. 2008).
 
We have applied the same procedure also to the VVDS parent samples described in Section~\ref{selcrit}:
all available photometry has been used for each galaxy, together with
spectroscopy, renormalized to the {\it i'}-band magnitude, or to the $I$-band 
magntitude when the previous one was not available (as it is the case
for the VVDS-14h field). As the parent sample comprises all kinds of
galaxies, we have not imposed a prior on the $D_{4000}$ break.

Stellar mass and star formation rate derived from the SED fitting (median values 
of the PDF and associated uncertainties) for MASSIV galaxies are listed in 
Table~\ref{massivsample}. 
	
The stellar mass distributions of the MASSIV sample are shown in Figure~\ref{masshist}. Median values and associated dispersion are listed in Table~\ref{table:2}. MASSIV galaxies are distributed in the stellar mass range between $\sim 3 \times 10^{9}$ and $6 \times 10^{11}$ \msun. The median stellar mass for the MASSIV sample is $1.4 \times 10^{10}$ \msun. The median stellar masses for the MASSIV galaxies drawn from the Wide, Deep and Ultra-Deep VVDS samples are $4.8 \times 10^{10}$ \msun,  $1.2 \times 10^{10}$ \msun, and $1.4 \times 10^{10}$ \msun\ respectively.
				
The SED-derived star formation rate distributions of the MASSIV sample are shown in Figure~\ref{sfrsedhist}. Median values and associated dispersion are listed in Table~\ref{table:2}. MASSIV galaxies are distributed in the SFR range between $\sim 5$ to $400$ \msunpyr. The median SED-derived SFR for the MASSIV sample is $\sim 32$ \msunpyr. The median star formation rates for the MASSIV galaxies drawn from the Wide, Deep and Ultra-Deep VVDS samples are $\sim 63$, $23$, and $31$  \msunpyr\ respectively.

\begin{figure}[t]
   \centering
   \includegraphics[bb=18 144 592 718,width=9cm]{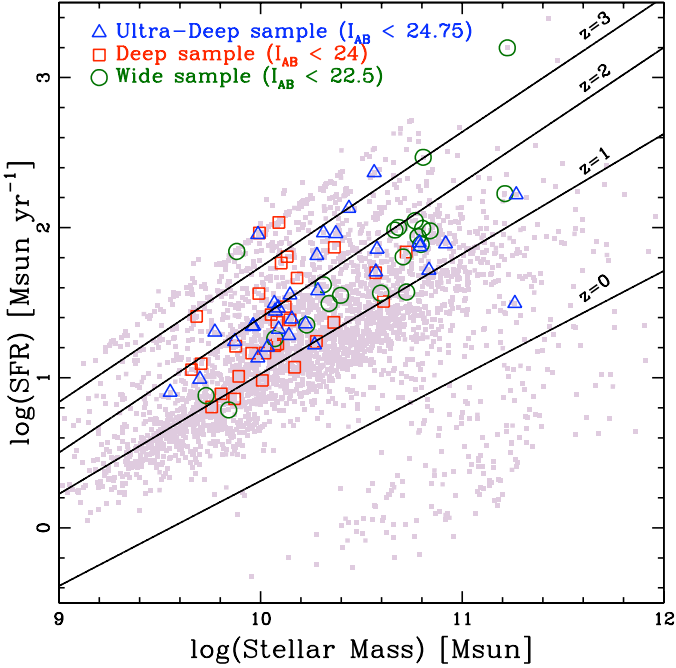}
      \caption{
      SED-derived star formation rate as a function of stellar mass. The MASSIV sample (Wide = green circles, Deep = red squares  and Ultra-Deep = blue triangles) is compared 
      to the VVDS parent sample (small filled squares). The lines represent the empirical relations between SFR and stellar mass for different redshifts between $z=0$ and $z=3$ following the analytical expression given in Bouch\'e et al. (2010).
                          }
         \label{sfrsedvsmass}
   \end{figure}

The relation between the SED-based SFR and stellar mass is shown in Figure~\ref{sfrsedvsmass}. The location of MASSIV galaxies selected in the Wide, Deep and Ultra-Deep catalogues is compared to the one of the parent VVDS sample. Most of the MASSIV galaxies occupy an area bounded by the star formation ``main sequences" defined empirically 
at $z\sim 1$ and $z\sim 2$ (Bouch\'e et al. 2010). This figure shows cleary that the MASSIV sample is representative of the overall population of star-forming galaxies at $z\sim 1-2$. 

	\subsection{Discussion on selection effects}
	\label{seleffects}

Having been assembled using different flux-limited VVDS catalogues, it is worth assessing what part of the $z\sim 1-2$ 
population is represented by our MASSIV sample with respect to the ÒunbiasedÓ population of high-redshift galaxies probed by the VVDS. Indeed, VVDS is a purely apparent magnitude-selected sample, without any color selection unlike color-selected samples produced in the same redshift range as MASSIV (e.g. BzK, BM/BX). 
The VVDS sample is thus representative of the overall star-forming population of galaxies at high redshifts. 

The most stringent selection criteria used to built the MASSIV sample is certainly the requirement of a minimum \oii\ equivalent width  (see Sect.~\ref{selcrit}). This sensitivity limit is likely to translate into an overall ``bias" towards younger and more actively star-forming systems. To check for this possible effect, we compare in Figure~\ref{selection} the \oii\ flux and EW distributions of the final MASSIV sample to those derived for the VVDS parent catalogues. 
We clearly see that the selection criteria based on the \oii\ EW ends up with  
a sample of parent VVDS galaxies (empty-solid histograms in Fig.~\ref{selection}) with higher median values of the \oii\ EW ($-54\AA$ compared with $-38\AA$) and flux ($6.9$ compared with $4.4 \times 10^{-17}$ \ergscm).  However, the final random selection of MASSIV targets (filled cyan histograms in Fig.~\ref{selection}) among this parent sample does not introduce any additional strong selection effect as the median values of \oii\ EW and flux ($-61\AA$ and $8.3\times 10^{-17}$ \ergscm) do not vary significantly, even if we slightly tend to select the strongest \oii\ emitters. A Kolmogorov-Smirnov (K-S) statistical test performed on the \oii\ flux distributions of the MASSIV and [OII]-selected VVDS samples further shows a high probability for these two samples to be drawn from the same distribution.

With a selection based on \oii\ emission-line or UV intensity, MASSIV is thus essentially a flux-limited sample, translating into a SFR lower limit for a given stellar mass (see Fig.~\ref{sfrsedvsmass}), and reaching SFRs as low as a few \msunpyr.

Looking at the stellar mass distributions in Figure~\ref{masshist} and median values listed in Table~\ref{table:2}, we see clearly that we are probing different mass regimes depending on the  considered VVDS parent sample. The Wide VVDS sample, with its limiting magnitude \iab\ $\leq 22.5$, contains more massive galaxies than the Deep/Ultra-Deep samples (\iab\ $\leq 24$ and $24.75$ respectively). This is due both to the larger area and shallower limiting magnitude of the wide survey compared with the deeper ones. But the important point is that the MASSIV sample, drawn from the Wide, Deep, and Ultra-Deep VVDS samples, is probing a large and representative range of stellar masses allowing us, in fine, to 
draw robust conclusions on the nature of the overall population at high redshifts. The part of the VVDS  $z\sim 1-2$  population that is most clearly absent among the MASSIV sample is the massive quiescent tail at low star formation rates (see Figures~\ref{masshist} and \ref{sfrsedhist}).  Indeed such objects would be difficult to detect as no or very faint \oii\ is expected, at least from star formation.
	
We conclude from this section that in spite of the selection criteria (from the VVDS parent survey/catalogues
and the additional specific criteria considered in choosing our targets), the MASSIV sample provides a good representation of ``normal" star-forming galaxies, with a median SFR $\sim 30$ \msunpyr, in the stellar mass regime $10^9 - 10^{11}$ \msun.  

\section{Composite integrated spectra}
\label{spectra}

   \begin{figure}[t]
   \centering
   \includegraphics[bb=18 144 592 718,width=9cm]{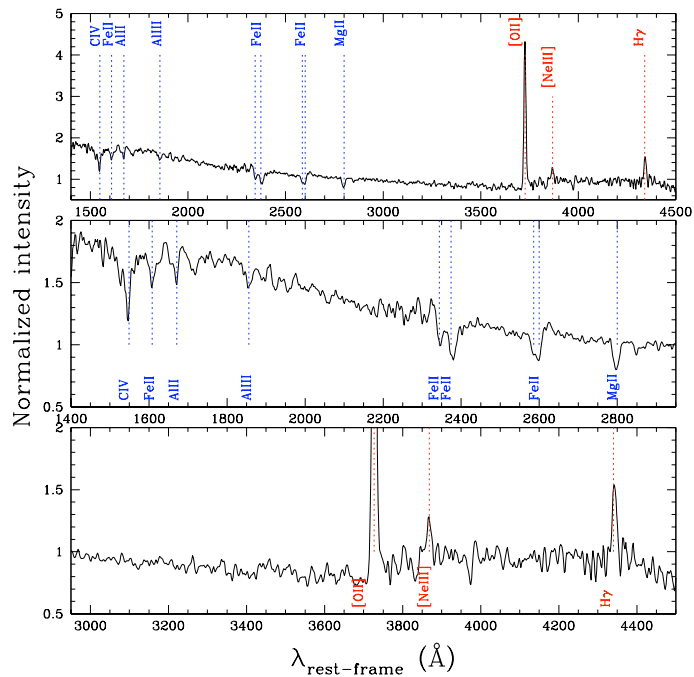}
      \caption{{\it Top}: Rest-frame stacked VIMOS spectrum for the full MASSIV sample (83 galaxies after exclusion of VVDS220148046 at $z=2.24$) normalized at the continuum level at $\lambda = 3000$\AA. The position of main absorption (blue) and emission (red) lines is indicated with dotted lines. The {\it middle} and {\it bottom} panels are for the same spectrum but zoomed-in on specific wavelength ranges.}
         \label{vimoscomposite}
   \end{figure}

   \begin{figure}[t]
  \centering
   \includegraphics[bb=18 144 592 718,width=9cm]{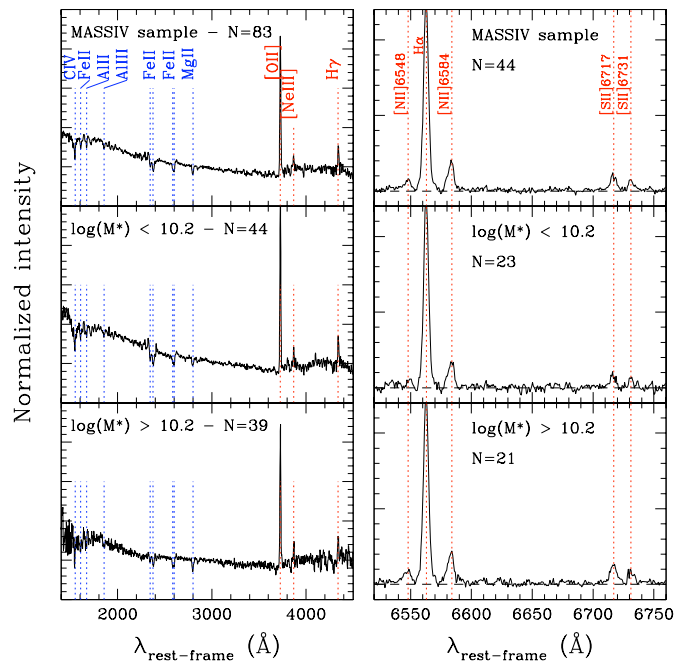}
      \caption{Rest-frame composite VIMOS (left) and SINFONI (right) 1D spectra,  normalized at the continuum level at $\lambda = 3000$\AA\ (VIMOS) or at the \halpha\ intensity level (SINFONI). Top panels: spectra are shown for the full (left) and ``first epoch" (right) MASSIV samples. Spectra corresponding  to two stellar mass bins are shown in middle ($ < 1.6 \times 10^{10}$ \msun) and bottom ($ > 1.6 \times 10^{10}$ \msun) panels.  The position of main absorption (blue) and emission (red) lines are indicated with dotted lines. In each panel, the number of galaxies used to build the composite spectrum is indicated.}
         \label{vimossinfonicomposite}
   \end{figure}

	\subsection{VIMOS spectra}
	
Composite VIMOS spectra have been built from integrated 1D spectra of MASSIV galaxies extracted from the VVDS database. Each spectrum has been first brought to rest frame (using the VVDS redshift) and then normalized to unity over the interval $2975-3025$\AA. The result of the averaging procedure is shown in Figure~\ref{vimoscomposite}, which presents the composite spectrum for the 83 MASSIV galaxies\footnote{A similar composite spectrum has been built using the 44 MASSIV galaxies of the first epoch sample only. No significant difference has been found from that for the full sample.} (the galaxy VVDS220148046 has been excluded due to its redshift $z=2.24$) in the rest-frame wavelength range $\sim 1500-4500$\AA. The high S/N of this composite spectrum ($\sim 30$ or typically $5-10$ times that of individual spectra in the wavelength range $2400-3800$\AA) is apparent. Compared to individual spectra, we see significantly more absorption lines (indicated with blue labels) in the composite as well as faint emission lines (\neiii\ and \hgamma) barely detected in individual frames. Composite spectra have also been produced in two different stellar mass bins, below (44 galaxies) and above (39 galaxies) a stellar mass of $1.6 \times 10^{10}$ \msun, corresponding to the median value of the MASSIV sample. These composite spectra are shown in Figure~\ref{vimossinfonicomposite} and are used in the next sections to investigate any mass dependence on the AGN contamination and integrated metallicity. The emission line equivalent widths measured in the three composite spectra are listed in Table~\ref{table:3}.

	\subsection{SINFONI spectra}

Composite SINFONI spectra have been built from integrated 1D spectra of the ``first epoch" sample of  MASSIV galaxies. Each spectrum has been obtained following the procedure detailed in Queyrel et al. (2011), corrected for the ``artifical" line broadening due to large-scale motions, and brought to rest frame using the SINFONI redshift.  Spectra are then normalized in order to set the \halpha\ peak intensity 
to unity. The result of the averaging procedure is shown in Figure~\ref{vimossinfonicomposite}, which presents the composite spectrum for 44 MASSIV galaxies\footnote{44 galaxies instead of 50 because 4 objects are targeted in \oiiib, including the one at $\sim 2.24$, and 2 galaxies were not detected} in the rest-frame wavelength range $\sim 6450-6750$\AA. The high S/N of this composite spectrum allows one to see clearly the \nii\ an \sii\ emission lines which are barely detected in a high fraction of individual frames (see Queyrel et al. 2011). Composite spectra have been also produced in two different stellar mass bins, below (23 galaxies) and above (21 galaxies) the median value ($1.6 \times 10^{10}$ \msun) of stellar mass. These composite spectra are shown in Figure~\ref{vimossinfonicomposite}. The emission line flux ratios measured in the three composite spectra are listed in Table~\ref{table:3}.

	\subsection{Physical properties derived from composite spectra}
		
	Thanks to their high S/N, the VIMOS and SINFONI composite spectra can be used to measure faint emission lines, such as \neiii, \hgamma, \nii, or \sii, which are 
	barely detected in a few individual galaxies only. These emission lines can then be used to probe some average physical properties of the MASSIV sample, such as the origin of the ionizing photons (starburst vs. AGN)  and their global metallicity. 
	
   \begin{table*}
   \tabcolsep1.40mm
\caption{Emission-line equivalent widths and ratios measured in VIMOS and SINFONI composite spectra}             
\label{table:3}      
\centering                          
\begin{tabular}{c c c c c c c }        
\hline\hline                 
MASSIV Sample & EW(\oii) & EW(\neiii) & EW(\hgamma) & \niib/\halpha\ & \sii/\halpha\ & \siir\ \\    
 & [\AA] &  [\AA] &  [\AA] & & & \\
\hline                        
Total & $-53.3\pm2.6$ & $-5.4\pm1.1$ & $-6.7\pm1.5$ & $0.154\pm0.012$ & $0.137\pm0.010$ & $1.54\pm0.11$ \\
log($M_{\rm *}$ [\msun]) $\leq 10.2$ & $-61.2\pm3.1$ & $-6.8\pm1.2$ & $-6.6\pm1.3$ & $0.140\pm0.016$ & $0.112\pm0.015$ & $1.44\pm0.16$ \\
log($M_{\rm *}$ [\msun]) $> 10.2$ & $-45.9\pm2.8$ & $-3.6\pm1.2$ & $-6.4\pm1.2$ & $0.170\pm0.015$ & $0.178\pm0.016$ & $1.64\pm0.15$ \\
\hline                                   
\end{tabular}
\end{table*}

	\subsubsection{AGN contamination?}
	
	MASSIV targets have been selected to be star-forming galaxies. However, even if type-1 AGNs  are clearly identified in the VVDS sample thanks to their broad emission 
	lines, the MASSIV sample could include a non-negligible fraction of type-2 AGNs (typically a few percent). We can use the composite spectra to check if low-activity type-2 AGNs are present in our sample, 
	by measuring typical emission-line ratios commonly used to distinguish star-forming galaxies from AGNs. 
	
\begin{figure}
  \centering
   \includegraphics[bb=18 144 592 718,width=4.4cm]{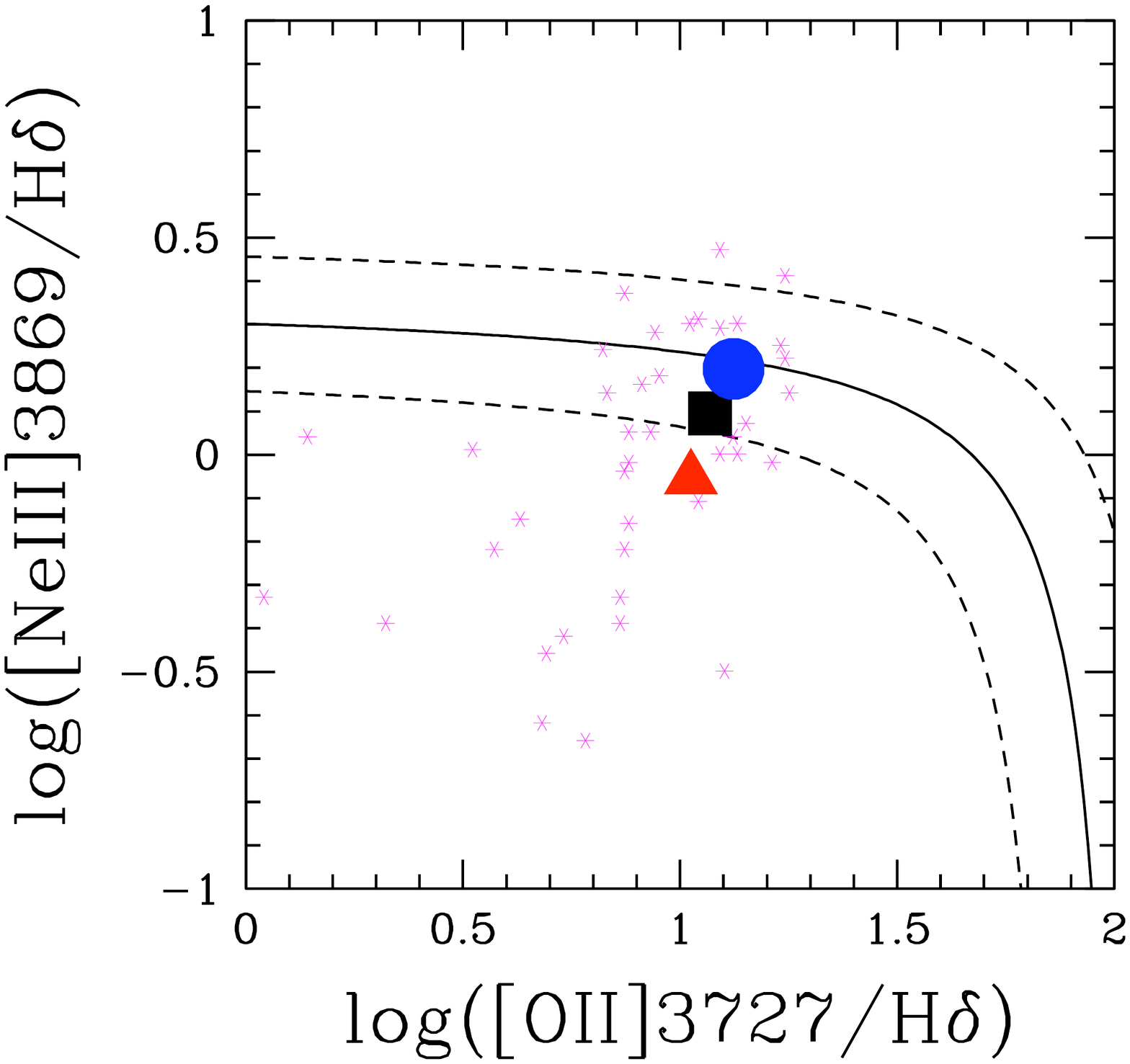}
   \includegraphics[bb=18 144 592 718,width=4.4cm]{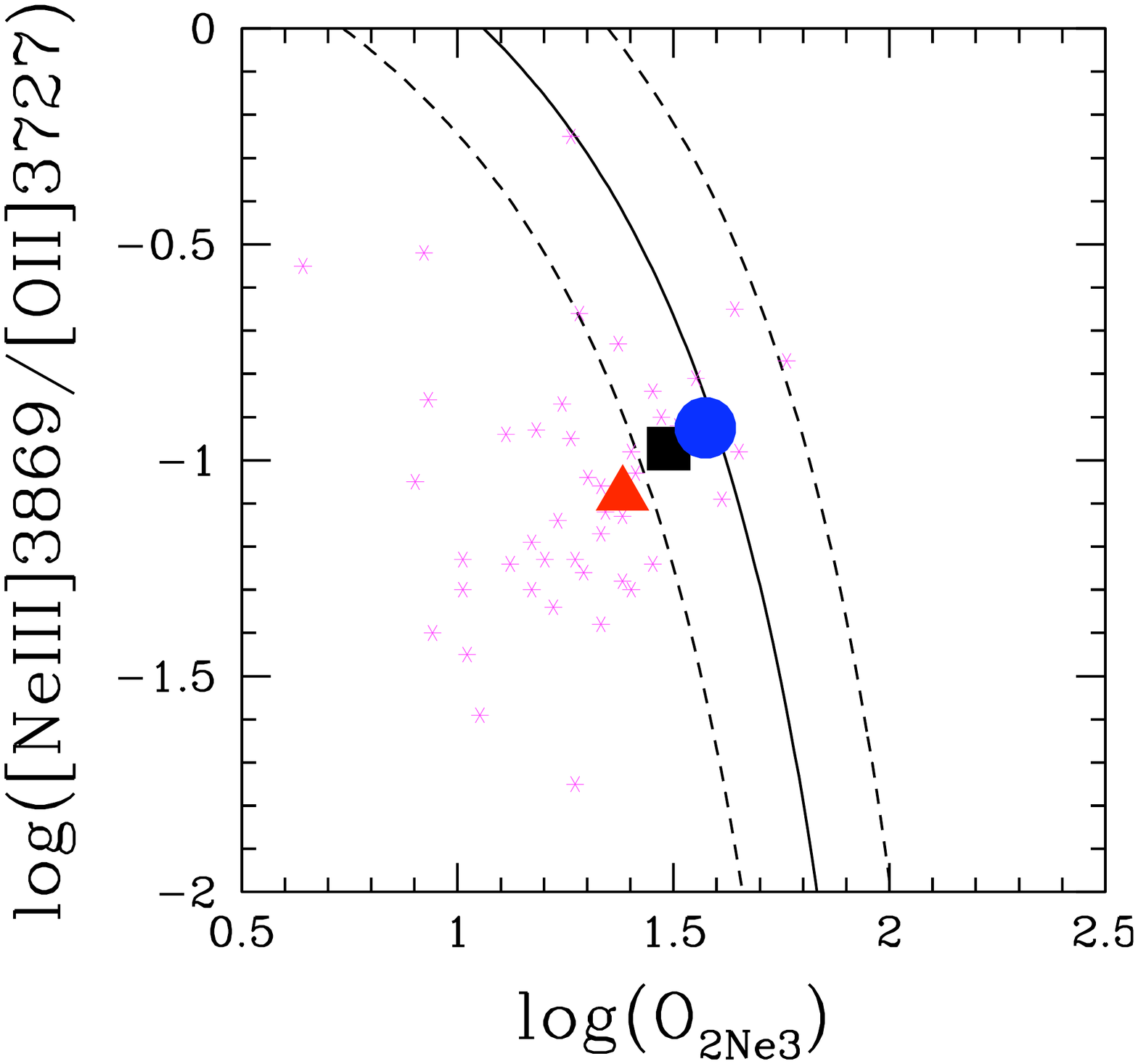}
      \caption{Diagnostic diagrams based on \neiii\ emission line. In the left-hand panel, the relation between the emission-line ratios \neiii/\hdelta\ and \oii/\hdelta\  is shown and in 
      the right-hand panel the relation between the emission-line ratio \neiii/\oii\ and the  $O_{\rm 2Ne3}$=$($\oii$+15.37$\neiii)/\hdelta\ parameter is shown. The solid lines represent the analytical division 
      (see Perez-Montero et al. 2007) between star-forming galaxies (lower left) and AGNs (upper right).   The dashed lines represent the limit of the bands of uncertainty of 0.15 dex on
both sides of each relation. Black square, blue point, and red triangle show the location of emission-line ratios measured in the composite spectrum of the full MASSIV sample, of galaxies with stellar mass lower than $1.6 \times 10^{10}$ \msun, and of galaxies with stellar mass higher than $1.6 \times 10^{10}$ \msun, respectively. Magenta asterisks show the location of individual VVDS galaxies with measured \neiii\ emission line (Perez-Montero et al. 2009).}
         \label{specdiag}
   \end{figure}

	Using the emission lines measured in the VIMOS composite spectrum, namely \oii, \neiii, and \hgamma\ (see Table~\ref{table:3}), we can use the line ratios based on the \neiii\ emission line and introduced by Perez-Montero et al. (2007) as a first 
	check for a possible contamination by narrow-line AGNs. These ratios involve the \hdelta\ emission line (see Figure~\ref{specdiag}) which can be deduced from the \hgamma\ emission line assuming a Balmer line ratio \hgamma/\hdelta\ $=1.82$ typical of star-forming galaxies (Osterbrock 1989). The location in the diagnostic diagrams of emission-line 
	ratios measured in the three VIMOS composite spectra (i.e. for the full MASSIV sample and for the two stellar mass bins) are shown in Figure~\ref{specdiag}, together 
	with the empirical limits (and associated uncertainty regions; see Perez-Montero et al. 2007) between  star-forming objects and AGNs.	Emission-line ratios for the 
	full MASSIV sample and for the two stellar mass bins are both very close to the separation limit between star-forming galaxies and AGNs. Note, however, that the emission-line ratios which are the closest  
	to this limit are those measured in the composite spectrum of the low-mass ($< 1.6 \times 10^{10}$ \msun) galaxies contrary to what would be expected if the  
	MASSIV sample was contaminated by a significant fraction of narrow-line AGNs, as active nuclei are preferentially located in the center of massive galaxies. The relatively high intensity of \neiii\ emission-line in low-mass MASSIV galaxies  is certainly due to their low metallicity (see section~\ref{metallicity}).  MASSIV galaxies populating the high stellar mass range show emission-line ratio typical of star-forming galaxies, even taking into account the uncertainty region. Their low-intensity \neiii\ emission-line can be explained by their higher metallicity (see section~\ref{metallicity}).
	
	The SINFONI composite spectra can also be used to check for a possible narrow-line AGN contamination in the MASSIV sample. Line ratios involving the \niib\ and \sii\ emission lines are listed in Table~\ref{table:3} for the three composite spectra. Unfortunately, we do not have access to the \oiiib/\hbeta\ line ratios commonly used in standard diagnostic diagrams. However, the \niib/\halpha\ and \sii/\halpha\ line ratios measured in the MASSIV composite spectrum ($\sim0.15$ and $0.14$ respectively) are low enough  to secure the MASSIV galaxies in the star-forming region of the diagnostic diagrams (see e.g. Lamareille et al. 2006). The difference in the measured line ratios between low-mass and massive galaxies is probably due also to a metallicity effect. 
	
	Based on these arguments, we thus conclude for a low contamination ($\leq$ 2--3\%) by type-2 AGNs, if any, in the MASSIV sample.
		
		\subsubsection{Global metallicity}
		\label{metallicity}
		
An estimate of the average gas-phase oxygen abundance in the MASSIV sample can be obtained from the emission-line ratios measured in the composite spectra. We refer to Queyrel et al. (2011) for a detailed study of spatially-resolved metallicity in the individual MASSIV galaxies. 

Using the VIMOS composite spectra, the oxygen abundance has 
been estimated from line ratios involving \oii, \neiii, and \hdelta\ emission lines  together with the calibrations proposed by Perez-Montero et al. (2007). This method has been already used to derive the mass-metallicity relation of VVDS galaxies in the redshift range $0.9 \leq z \leq 1.24$ (Perez-Montero et al. 2009). We find an oxygen abundance 12+log(O/H) $= 8.47$, $8.32$, and $8.62$ for the full MASSIV sample,  for galaxies with stellar mass lower than $1.6 \times 10^{10}$ \msun, and for galaxies with stellar mass higher than $1.6 \times 10^{10}$ \msun, respectively. 

Some authors (eg. Maier et al. 2006; Liu et al. 2008) claim that the typical physical properties (electron density, hardness of the radiation field, etc) of the ionized gas in high-redshift star-forming galaxies may be different from the ones observed in the local universe, which would affect the validity of the metallicity calibrations based on nearby galaxy samples. Whereas it is not possible to check for the ionization conditions through the \oiiisoii\  line ratio in our MASSIV sample as we are missing the \oiiib\ emission-line, we can make use of the measured \siissii\ line ratios listed in Table~\ref{table:3} to infer the average electron density (Osterbrock 1989).  Taking into account the measurement uncertainties, we find electron densities of the order of $\sim100$ \cmc\ (independently of the stellar mass range), which are typical of nebular conditions in local star-forming galaxies. We are thus quite confident in using metallicity calibrations based on local samples. 

Another independent estimate of the global metallicity of MASSIV galaxies is obtained from the \niib/\halpha\ line ratio measured in the SINFONI composite spectra. The calibration proposed by Perez-Montero \& Contini (2009) is used as in Queyrel et al. (2011).  We find an oxygen abundance 12+log(O/H) $= 8.43$, $8.39$, and $8.46$ for the full MASSIV sample,  for galaxies with stellar mass lower than $1.6 \times 10^{10}$ \msun, and for galaxies with stellar mass higher than $1.6 \times 10^{10}$ \msun, respectively. The global metallicity of MASSIV galaxies estimated from two different datasets and calibrations are very consistent and show, as expected from 
the mass-metallicity relation, a higher value for massive galaxies.  

\section{Comparison with other IFU samples of high-redshift galaxies}
\label{comparison}

For the sake of consistency and to allow for an homogeneous and fair comparison between the various galaxy samples described below, we refer to an unique IMF (Salpeter 1955) for the derivation of stellar masses and star formation rates. Following Bernardi et al. (2010), we thus applied a constant scaling factor of $+0.25$ or $+0.15$ to those quantities derived using a Chabrier (2003; for SINS, OSIRIS and LSD/AMAZE samples) or a ``diet" Salpeter (for IMAGES sample) IMF in the SED fitting procedure.

	\subsection{IMAGES}
	
IMAGES (Intermediate MAss Galaxies Evolution Sequence) survey targeted galaxies in the Chandra Deep Field South, with redshifts $z\sim 0.4-0.75$, \iab\ $\leq 23.5$ and detected \oii\ emission lines with 
rest-frame EW $\leq -15$\AA\ (Ravikumar et al. 2007). Galaxies were further selected to be of intermediate mass, using $J$-band absolute magnitudes as a proxy 
for stellar mass, such that $M_J \leq -20.3$, which roughly corresponds to \mstar\ $\geq 1.5 \times 10^{10}$ \msun. The IMAGES galaxies, which are 
representative of the $J$-band luminosity function at these redshifts (Yang et al. 2008), have been observed with the multi-integral-field unit spectrograph 
FLAMES/GIRAFFE (with a sampling of 0.52\arcsec/pixel) at the VLT

The final IMAGES sample contains a total of 63 galaxies with IFU resolved kinematics. The redshift range of these galaxies extends from $z\sim 0.4$ to $z\sim 0.75$ with a median 
value of $\sim 0.61$ (see Figs.~\ref{zhist_complitt} and \ref{zhist_complitt_images}). The stellar masses have been estimated for 40 IMAGES galaxies (Puech et al. 2010; Neichel et al. 2008) using a ``diet" Salpeter IMF and Bell et al. (2003) simplified recipes for deriving stellar mass from $J$-band luminosity.  The distribution of stellar masses for the IMAGES sample is shown in Fig.~\ref{masshist_complitt}. Stellar masses range from $\sim 6.3 \times 10^9$ to $2 \times 10^{11}$ \msun\ with a median value of $3.5 \times 10^{10}$ \msun. There is no estimate of 
the SED-based SFR for the IMAGES sample. However, the ``total" SFR has been computed (see Puech et al. 2007 for details) as the sum of SFR$_{\rm UV}$, derived from the 
2800\AA\ luminosity, and SFR$_{\rm IR}$, derived from Spitzer/MIPS photometry at 24\micron\ using the Chary \& Elbaz (2001) calibration between rest-frame 15\micron\ flux 
and the total IR luminosity. The calibrations of Kennicutt (1998), which rely on a Salpeter IMF, have been used to convert both UV and IR luminosities into SFRs. The distribution of 
total SFR for 36 IMAGES galaxies is shown in Fig.~\ref{sfrhist_complitt}. Total SFR ranges from $\sim 5$ to $80$ \msunpyr, with a median value of $10$ \msunpyr.

	\subsection{SINS}
	
With a total of 62 galaxies at $1.4 < z < 2.6$ detected in \halpha, the SINS (Spectroscopic Imaging survey in the Near-infrared with SINFONI; F{\"o}rster Schreiber et al. 2009) survey, carried out with SINFONI at the VLT, is one of 
the largest surveys of near-IR integral field spectroscopy to date. 
The SINS galaxies were selected from the spectroscopically-confirmed subsets of various imaging surveys in the 
optical, near-IR, and mid-IR. The photometric selection of the parent samples encompassed a range of star-forming populations at high redshift, including optically-selected ``BX/BM" galaxies at $z\sim 2-3$, and infrared-selected galaxies at  $z\sim 1.5-2.5$, with a majority of ``BzK" objects. The SINS targets were further selected to have an integrated emission line (\halpha\ or \oiiib\ in a few cases) flux of at least $5 \times 10^{-17}$ \ergscm. Although with some bias towards the bluer part of the galaxy population compared to purely $K$-band selected samples at similar redshifts, the SINS \halpha\ sample provides a reasonable representation of massive actively star-forming galaxies at $z\sim 2$.

\begin{figure}[t]
   \centering
   \includegraphics[bb=18 144 592 718,width=9cm]{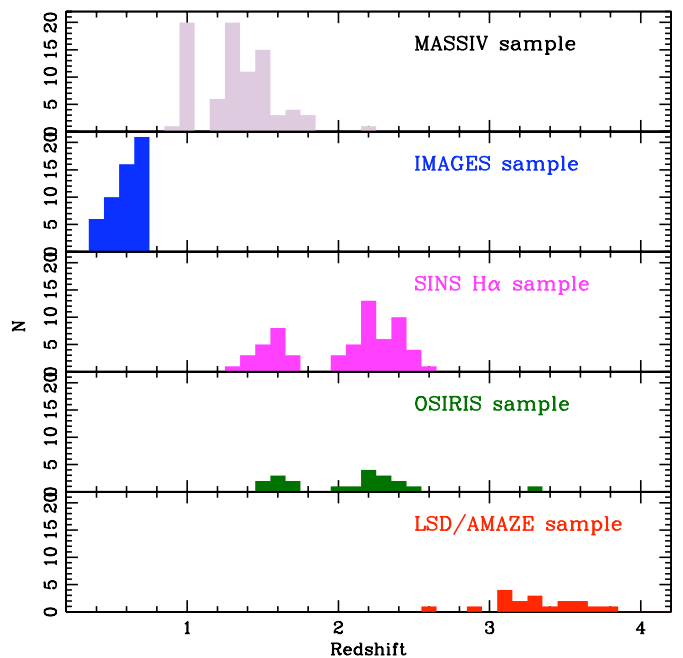}
      \caption{Redshift distribution of the MASSIV sample (top panel) compared with other samples of high-z galaxies observed with IFUs: the IMAGES sample (2nd panel from the top), the SINS  \halpha\ sample (3rd panel from the top), the OSIRIS sample (4th panel from the top), and the LSD/AMAZE sample (bottom panel). 
                          }
         \label{zhist_complitt_images}
   \end{figure}

\begin{figure}[t]
   \centering
   \includegraphics[width=10.5cm]{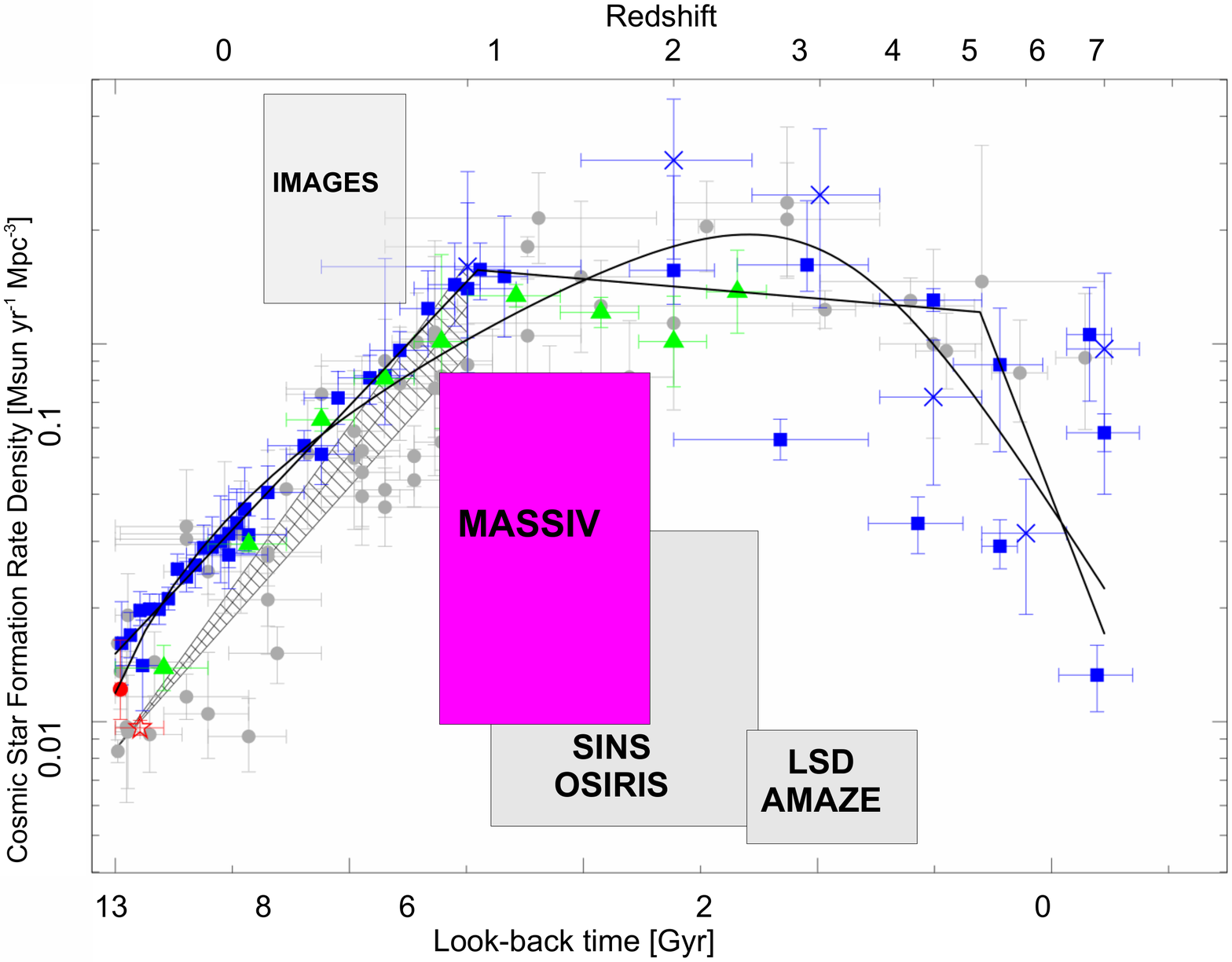}
      \caption{Evolution of the cosmic star formation rate density as a function of look-back time and redshift (adapted from Hopkins 2006). The redshift range of MASSIV ($0.9 < z < 1.8$, magenta box) is compared with other major IFU surveys of distant galaxies: IMAGES ($z \sim 0.4-0.75$), SINS/OSIRIS ($z \sim 1.4-2.6$), and LSD/AMAZE ($z \sim 2.6-3.8$) . The relative height of each boxes is proportional to the samples size.
                                }
         \label{zhist_complitt}
   \end{figure}

\begin{figure}[t]
   \centering
   \includegraphics[bb=18 144 592 718,width=9cm]{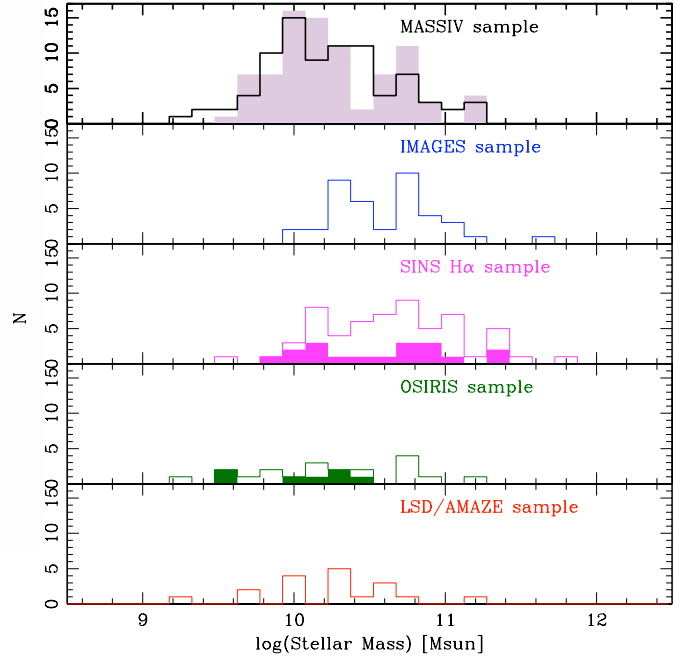}
      \caption{Stellar mass distribution of the MASSIV sample (top panel). The filled (empty) histogram is for PDF (``best fit") masses. This distribution is compared with other samples of high-z galaxies observed with IFUs: the IMAGES sample (2nd panel from the top), the SINS \halpha\ sample (3rd panel from the top), the OSIRIS sample (4th panel from the top), and the LSD/AMAZE sample (bottom panel). The filled histograms in the 3rd and 4th panel are restricted to $z < 2$ galaxies. For comparison between the different samples, a scaling factor of $+0.25$ dex (SINS, OSIRIS, and LSD/AMAZE samples) or $+0.15$ dex (IMAGES sample) has been applied to the published stellar masses.
                          }
         \label{masshist_complitt}
   \end{figure}

\begin{figure}[t]
   \centering
   \includegraphics[bb=18 144 592 718,width=9cm]{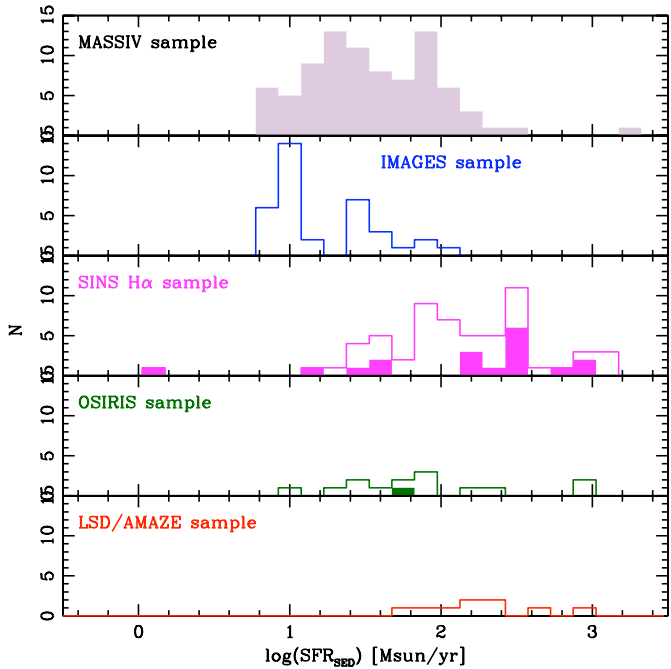}
      \caption{SED-derived SFR distribution of the MASSIV sample (top panel) compared with other samples of high-z galaxies observed with IFUs: the IMAGES sample (2nd panel from the top), the SINS  \halpha\ sample (3rd panel from the top), the OSIRIS sample (4th panel from the top), and the LSD/AMAZE sample (bottom panel). The filled histograms are restricted to $z < 2$ galaxies. For comparison between the different samples, a scaling factor of $+0.25$ dex (SINS, OSIRIS, and LSD/AMAZE samples) or $+0.15$ dex (IMAGES sample) has been applied to the published star formation rates.
                          }
         \label{sfrhist_complitt}
   \end{figure}

The redshift histogram of the SINS galaxies (see Fig.~\ref{zhist_complitt_images}) shows a bimodal distribution, peaking at $z\sim 1.6$ and $z\sim 2.3$, and has a median value of $z\sim 2.17$. The reason for the bimodal distribution is simply the result of the requirement that the emission line of interest (primarily \halpha) falls within either the $H$ or $K$ band atmospheric windows. 
The stellar masses have been estimated for 59 SINS galaxies using a standard SED fitting procedure based on extensive photometry (see details in F{\"o}rster Schreiber et al. 2009), and Bruzual \& Charlot (2003) stellar population synthesis models, with a constant SFR, solar metallicity, and a Chabrier (2003) IMF.  The distribution of stellar masses for the SINS sample is shown in Figure~\ref{masshist_complitt}. Stellar masses range from $\sim 3 \times 10^9$ to $5 \times 10^{11}$ \msun\ with a median value of $4.6 \times 10^{10}$ \msun. SED-based star formation rates have also been derived from stellar population models for this SINS sub-sample of 59 galaxies. Their distribution is shown in Figure~\ref{sfrhist_complitt}. SED-based SFR ranges from $\sim 1$ to $1300$ \msunpyr, with a median value of $129$ \msunpyr.

   \begin{figure}[t]
   \centering
   \includegraphics[bb=18 144 592 718,width=9cm]{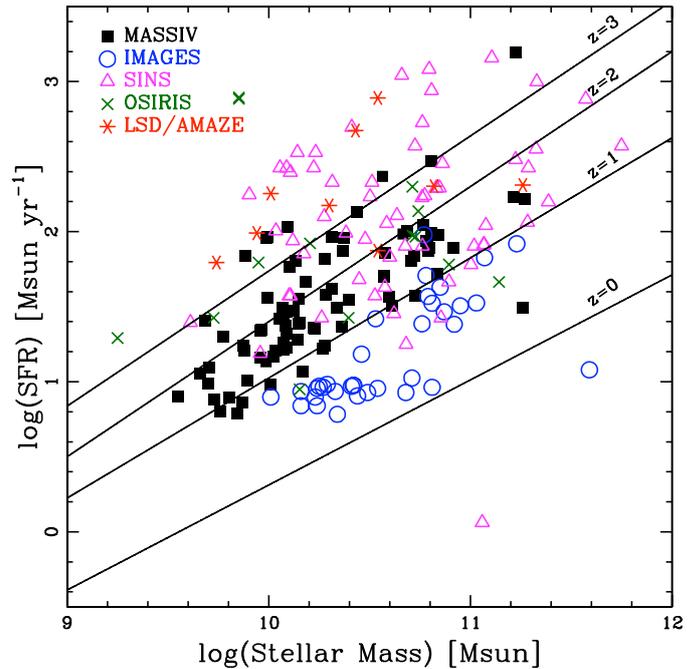}
      \caption{
      SED-derived star formation rate as a function of stellar mass. The MASSIV sample (black filled squares) is compared with other major IFU surveys of distant galaxies: IMAGES ($z \sim 0.4-0.75$, blue circles), SINS ($z \sim 1.4-2.6$, magenta triangles), OSIRIS ($z \sim 1.5-3.3$, green crosses), and LSD/AMAZE ($z \sim 2.6-3.8$, red asterisks). The lines represent the empirical relations between SFR and stellar mass for different redshifts between $z=0$ and $z=3$ following the analytical expression given in Bouch\'e et al. (2010). For comparison between the different samples, a scaling factor of $+0.25$ dex (SINS, OSIRIS, and LSD/AMAZE samples) or $+0.15$ dex (IMAGES sample) has been applied to the published stellar masses and star formation rates.
                          }
         \label{sfrsedvsmass_compalitt}
   \end{figure}
	
   \begin{table*}
\caption{The MASSIV sample compared with other major IFU surveys of high-redshift galaxies}             
\label{table:4}      
\centering                          
\begin{tabular}{l c r c l r l r }        
\hline\hline                 
Sample &  \multicolumn{3}{c}{Redshift} & \multicolumn{2}{c}{log(Stellar Mass) [\msun]} & \multicolumn{2}{c}{log(SFR$_{\rm SED}$)$^1$  [\msunpyr]}  \\    
 & Range  & $Median\pm \sigma$ & $N$ &  $Median\pm \sigma$ & $N$ &  $Median\pm \sigma$ & $N$  \\
\hline                        
MASSIV & $0.9-1.8$ & $1.33\pm0.13$ & 83 & $10.15\pm 0.30$$^2$ & 83 & $1.50\pm 0.31$ & 83  \\
&  &  & & $10.21\pm 0.27$$^3$ & 83 &  &  \\
IMAGES & $0.4-0.75$ & $0.62\pm0.07$ & 63 & $10.54\pm 0.24$ & 40 & $0.98\pm 0.29$ & 36  \\
SINS &  $1.4-2.6$ & $2.17\pm0.36$ & 62 & $10.66\pm 0.37$ & 59 & $2.11\pm 0.31$ & 59  \\
&  $1.4-2.0$ &  & & $10.76\pm 0.38$$^4$ & 18 & $2.43\pm 0.48$$^4$ & 18  \\
OSIRIS &  $1.5-3.3$ & $2.17\pm0.32$ & 20 & $10.29\pm 0.43$ & 20 & $1.92\pm 0.35$ & 14  \\
&  $1.5-2.0$ & & &  $10.20\pm 0.41$$^5$ & 7 &  &  \\
LSD/AMAZE &  $2.6-3.8$ & $3.29\pm0.18$ & 18 & $10.31\pm 0.30$ & 18 & $2.25\pm 0.16$ & 9  \\
\hline                                   
\multicolumn{8}{l}{$^1$ Except for the IMAGES sample, see text for details}  \\
\multicolumn{8}{l}{$^2$ Stellar mass derived from SED fitting using the PDF method}  \\
\multicolumn{8}{l}{$^3$ Stellar mass derived from SED fitting using the ``best fit" method}  \\
\multicolumn{8}{l}{$^4$ Stellar mass and SFR derived from SED fitting for the SINS $z<2$ subsample only}  \\
\multicolumn{8}{l}{$^5$ Stellar mass derived from SED fitting for the OSIRIS $z<2$ subsample only}  \\
\end{tabular}
\end{table*}

	\subsection{OSIRIS sample}
	
Galaxies in the so-called ``OSIRIS  sample" (Law et al. 2007, 2009; Wright et al. 2007, 2009) have been selected from the rest-frame UV color-selected catalogue of Steidel et al. (2004), and are all confirmed with optical spectroscopy using LRIS on the Keck telescope. The Steidel et al. survey primarily focused on $z\sim 2-3$ galaxies, but has also used the two-color technique introduced in Adelberger et al. (2004) to detect lower redshift $z\sim 1.6$ star-forming counterparts within the traditional redshift desert (Steidel et al. 2004). The heterogeneous OSIRIS sample has been built following a variety of criteria. Some targets were deliberately selected for their young stellar population and correspondingly small stellar masses, some for their old ages and large stellar masses, and some for other reasons including complex or multi-component rest-frame UV morphologies (Law et al. 2007), strong detections in \halpha\ narrowband surveys, unusual spectral features, or previous acquisition of long-slit kinematic data. However, given the relatively shallow OSIRIS $K$-band sensitivity, the most common criterion applied was based on the emission line flux ($\geq 5 \times 10^{-17}$ \ergscm) previously measured with long-slit spectroscopy (Erb et al. 2006). 
IFU observations have been obtained at the Keck telescope using NIR integral-field spectrograph OSIRIS with the AO/LGS system. 

The OSIRIS sample contains a total of 20 galaxies with \halpha\ measurements. The redshift range of these galaxies extends from $z\sim 1.5$ to $z\sim 3.3$  (see Figs.~\ref{zhist_complitt} and \ref{zhist_complitt_images}), with a bimodal distribution peaking at $z\sim 1.6$ and $z\sim 2.2$, and a median value of $z\sim 2.17$.  
The reason for the bimodal distribution is the same as for the SINS sample: the emission line of interest (primarily \halpha) has to fall within either the $H$ or $K$ band atmospheric windows. The stellar masses have been estimated using a standard SED fitting procedure based on optical/NIR ($UGRJK$) and MIR/FIR (Spitzer IRAC and MIPS) photometry, and Bruzual \& Charlot (2003) stellar population synthesis models, with a constant SFR, solar metallicity, and a Chabrier (2003) IMF.  The distribution of stellar masses for the OSIRIS sample is shown in Figure~\ref{masshist_complitt}. Stellar masses range from $\sim 2 \times 10^9$ to $10^{11}$ \msun\ with a median value of $2.0 \times 10^{10}$ \msun. SED-based star formation rates have also been derived from stellar population models for a fraction (14 out of 20 galaxies) of the OSIRIS sample . Their distribution is shown in Figure~\ref{sfrhist_complitt}. SED-based SFR ranges from $\sim 8$ to $700$ \msunpyr, with a median value of $83$ \msunpyr.

	\subsection{AMAZE \& LSD}
	
AMAZE \& LSD are two surveys conducted with SINFONI, with the main aim of studying the metallicity and 
dynamics of high-redshift ($z \geq 3$) galaxies. 
The AMAZE (Assessing the Mass-Abundance redshift Evolution; Maiolino et al. 2008) sample consists of 30 Lyman Break Galaxies in the redshift range $3 < z < 4.8$ (most of which at $z\sim 3$), with deep Spitzer/IRAC photometry ($3.6-8$ \micron). These galaxies were observed with SINFONI in seeing-limited mode. 
LSD (Lyman-break galaxies Stellar populations and Dynamics; Mannucci et al. 2009) is a smaller sample of 10 LBGs at $z\sim 3$ with available Spitzer and HST imaging. For LSD, SINFONI observations were performed with the aid of adaptive optics in order to improve spatial resolution since this project was aimed to obtain spatially-resolved spectra for measuring kinematics and gradients in emission lines. 

The combined LSD/AMAZE sample (Gnerucci et al. 2011) contains a total of 35 galaxies with \oiiib\ emission line measurements, except for the nearest $z\sim2.6$ galaxy for which \halpha\ has been targeted. The redshift range of these galaxies extend from $z\sim 2.6$ to $z\sim 3.8$ with two galaxies only at $z\leq 3$  (see Fig.~\ref{zhist_complitt}), and a median value of $\sim 3.3$. The stellar masses have been estimated using a standard SED fitting procedure based on optical/NIR ($UGRJK$) and Spitzer IRAC photometry, and Bruzual \& Charlot (2003) stellar population synthesis models, with a smooth exponentially decreasing SFR, a $Z = 0.2$ \zsun\ metallicity, and a Chabrier (2003) IMF.  The distribution of stellar masses for the LSD/AMAZE sample is shown in Figure~\ref{masshist_complitt}. Stellar masses range from $\sim 2 \times 10^9$ to $10^{12}$ \msun\ with a median value of $2.0 \times 10^{10}$ \msun. SED-based star formation rates have also been derived from stellar population models for the AMAZE sample only (9 galaxies) . Their distribution is shown in Figure~\ref{sfrhist_complitt}. SED-based SFR ranges from $\sim 60$ to $750$ \msunpyr, with a median value of $178$ \msunpyr.

	\subsection{Comparison}
		
Table~\ref{table:4} summarizes the redshift range probed by the different IFU surveys presented in the previous sections, together with  the median values of redshift, stellar mass and SED-based SFR, allowing a direct comparison with MASSIV.
 
The redshift range probed with the MASSIV sample is compared with the other IFU samples in Figures~\ref{zhist_complitt_images} and~\ref{zhist_complitt}. MASSIV covers a redshift range ($z \sim 0.9-1.8$) which is intermediate between the high-z surveys ($z \sim 2-3$, SINS, OSIRIS and LSD/AMAZE) and the IMAGES sample ($z \sim 0.4-0.8$). However, the MASSIV, OSIRIS and SINS surveys are probing a common redshift range at $z \sim 1.3-1.8$, whereas LSD/AMAZE targets only galaxies at the highest redshifts  ($z \sim 2.6-3.8$). MASSIV could in principle follow the evolution of galaxy kinematics and mass assembly at a crucial epoch corresponding to a transition phase between non-relaxed systems observed at high redshifts and more evolved and stable galaxies which populate the Hubble sequence at low redshifts. 

The stellar mass distribution of the MASSIV galaxies is compared with the other IFU samples in Figure~\ref{masshist_complitt}. The adopted stellar masses for the MASSIV sample are those derived with the PDF method (see Section~\ref{sedfit}). However, in order to be able to compare with other surveys, we also derived the stellar masses with the so-called ``best fit" (BF hereafter) method which is the one adopted for the other IFU samples. Both stellar mass distributions (PDF and BF) are thus shown in Figure~\ref{masshist_complitt}. The stellar mass range probed by MASSIV is rather similar to the one covered by the IMAGES, SINS, OSIRIS, and LSD/AMAZE surveys, extending from a $\sim10^9$ to $10^{11}$ \msun. The median value of the MASSIV sample ($1.6 \times 10^{10}$ \msun\ with the BF method) is similar to the median values of the IMAGES/OSIRIS/LSD/AMAZE samples ($\sim2 \times 10^{10}$ \msun) but lower by a factor of $\sim3$ (at a confidence level $> 3\sigma$) than the median value of the SINS sample ($4.6 \times 10^{10}$ \msun).

The SED-based SFR distribution of the MASSIV sample is also compared with other IFU samples in Figure~\ref{sfrhist_complitt}. The lowest values of SFR probed by MASSIV (a few \msunpyr) is rather similar to the ones targeted by the other surveys, except for the highest redshift surveys LSD/AMAZE starting at $\sim 60$ \msunpyr. The median value of the MASSIV sample (32 \msunpyr) is, however, significantly lower (at a confidence level $> 3\sigma$) than the median values of the OSIRIS (83 \msunpyr), SINS (129 \msunpyr), and LSD/AMAZE (178 \msunpyr) samples.

In Figure~\ref{sfrsedvsmass_compalitt} we compare the location of the MASSIV galaxies to the other IFU samples in the SFR vs. stellar mass relationship. As already stated in sect.~\ref{sedfit}, the MASSIV sample is representative of the overall population of star-forming galaxies at $z\sim 1-2$. For a given stellar mass, the level of star formation probed by MASSIV is  globaly lower (at a confidence level $> 3\sigma$) that the one of SINS and LSD/AMAZE targets which reach easily SFRs higher than $\sim500$ \msunpyr. On the opposite, the IMAGES sample shows much lower SFRs ($\leq 100$ \msunpyr) for a given stellar mass as expected by the star formation ``main sequence" defined at $z\sim 0.4-0.8$ (see Bouch\'e et al. 2010).

Compared to other high-$z$ IFU surveys, the main advantage of MASSIV is its representativeness for star-forming galaxies with SFR $\geq5$ \msunpyr. Indeed, MASSIV has been drawn from VVDS, the most unbiased and representative sample of high-$z$ star-forming galaxies available so far. On the contrary, surveys like SINS, OSIRIS and LSD/AMAZE targeted galaxies selected in various (and hence heteregeneous) color-selected samples (LBG, BzK, sub-mm galaxies, etc), thus probing mainly galaxies at a relatively high star formation level (see Table~\ref{table:4}). 

One could argue that this difference in SFR is largely attributed to a combination of higher redshift ranges covered by other IFU surveys (except IMAGES) and the limits imposed on the minimum (expected) line flux or SFR to ensure detection. But if we restrict the SINS sample to the $1.4-2.0$ redshift range in common with MASSIV, the SINS galaxies have still a median SFR ($\sim270$ \msunpyr; 18 objects) significantly higher than the MASSIV one ($\sim60$ \msunpyr; 27 objects). The same is true for the expected lower limit on SFR imposed by a selection based on line flux (\oii\ for MASSIV and \halpha\ for a fraction of SINS galaxies). Using the quoted lower limit on \halpha\ flux used for the selection of SINS galaxies ($=5\times10^{-17}$ \ergscm;  F{\"o}rster Schreiber et al. 2009), we end up with a minimum SFR of $12$ \msunpyr, similar to the value derived for MASSIV ($8$ \msunpyr\ for \oii\ line flux $=1.5\times10^{-17}$ \ergscm), using the Argence \& Lamareille (2009) consistent calibrations to derive the SFR from line luminosities.

The total number of 84 galaxies included in MASSIV is also a clear advantage compared to surveys like OSIRIS and LSD/AMAZE which are limited to $\sim 20$ galaxies each.  The MASSIV sample indeed allows us to probe different mass and SFR ranges while keeping enough statistics in each category in order to derive robust conclusions on galaxy kinematics and mass assembly.

\section{Conclusions}
\label{conclusion}

This paper presents the basic physical properties of the full MASSIV sample containing 84 star-forming galaxies drawn from 
the VVDS in the redshift range $0.9 < z < 2.2$. SINFONI observations are now completed and the analysis of a first sample of 
50 MASSIV galaxies is presented in related papers focusing on the kinematical classification (Epinat et al. 2011), 
the evolution of fundamental scaling relations such as the baryonic Tully-Fischer one (Vergani et al. 2011), and the spatially-resolved metallicity (Queyrel et al. 2011).  
The selection criteria are mainly based on \oii\ emission line strength, making the MASSIV sample a fair representation of 
``normal" star-forming (with a median SFR $\sim 30$ \msunpyr) galaxies in the stellar mass regime $10^9 - 10^{11}$ \msun.  
Analysis based on composite VIMOS and SINFONI spectra reveals that the contamination by type-2 AGNs, if any, is very low ($\leq$ 2--3\%) and 
that the integrated metallicity of the galaxies follow the well-known mass-metallicity relation. Compared to other existing high-$z$ IFU surveys, 
we conclude that the main advantages of MASSIV are its representativeness (flux-selected from the magnitude-selected VVDS sample), 
and sample size allowing us to probe different mass and SFR ranges while keeping enough statistics in each 
category in order to derive robust conclusions on galaxy kinematics and mass assembly.

\begin{acknowledgements}
We warmly thank the anonymous referee for her(his) useful and constructive comments. This work has been partially supported by the
CNRS-INSU and its Programme National Cosmologie-Galaxies (France) and by the french ANR grant ANR-07-JCJC-0009. DV acknowledges the support from the INAF contract PRIN-2008/1.06.11.02. 
\end{acknowledgements}

\longtab{5}{
\tabcolsep1.40mm
\begin{longtable}{ccclcrrrrr}
\caption{\label{massivsample}Global properties of the MASSIV sample. Column 1 lists the VVDS identification number. Columns 2 and 3 give the RA and DEC coordinates. Columns 4  and 5 list the galaxy redshift and associated flag (see Le F\`evre et al. 2005) as measured in VIMOS spectra. Column 6 indicates the depth (D$=$Deep, W$=$Wide, UD$=$Ultra-Deep) of the VVDS parent sample. Column 7 gives  the galaxy stellar mass and associated uncertainty (in log of solar masses). Column 8 lists the galaxy SED-based star formation rate and associated error (in log of \msunpyr). Columns 9 and 10 indicate the \oii\ equivalent width (in $\AA$) and flux (in $10^{-17}$ \ergscm) and associated uncertainties as measured in VIMOS spectra}\\
\hline\hline
VVDS ID & RA(2000) & Dec(2000) & Redshift & $z_{\rm flag}$ & Sample & \mstar & SFR$_{\rm SED}$ & EW$_{\rm [OII]}$ & F$_{\rm [OII]}$ \\
(1) & (2) & (3) & (4) & (5) & (6) & (7) & (8) & (9) & (10) \\
\hline
\endfirsthead
\caption{continued.}\\
\hline\hline
VVDS ID & RA(2000) & Dec(2000) & Redshift & $z_{\rm flag}$ & Sample & \mstar & SFR$_{\rm SED}$ & EW$_{\rm [OII]}$ & F$_{\rm [OII]}$  \\
(1) & (2) & (3) & (4) & (5) & (6) & (7) & (8) & (9)  & (10) \\
\hline
\endhead
\hline
\endfoot

020106882&	02:25:21.800&	$-$4:46:18.602&	1.3980&	3& D&	 $9.99\pm0.22$&	$1.56\pm0.35$&	 $-61.0\pm 2.8 $&	$10.6\pm 0.6 $\\
020116027&	02:25:51.133&	$-$4:45:04.475&	1.5259&	2& D&	$10.09\pm0.23$&	$2.03\pm0.21$&	 $-57.0\pm 8.0$&	$6.0\pm 2.0$\\
020126402&	02:25:11.658&	$-$4:43:40.123&	1.2332&	2& D&	$10.09\pm0.26$&	$1.22\pm0.53$&	 $-60.0\pm 5.3 $&	$7.3\pm 0.8 $\\
020147106&	02:26:45.386&	$-$4:40:47.388&	1.5174&	2& D&	$10.10\pm0.38$&	$1.76\pm0.21$&	 $-78.0\pm 9.0$&	$9.0\pm 2.0$\\
020149061&	02:27:05.225&	$-$4:40:29.338&	1.2897&	3& D&	$10.18\pm0.23$&	$1.66\pm0.41$&	 $-45.9\pm 1.4 $&	$20.9\pm 0.7 $\\
020164388&	02:26:50.939&	$-$4:38:20.872&	1.3532&	3& D&	$10.13\pm0.31$&	$1.81\pm0.27$&	 $-112.4\pm 6.4 $&	$44.6\pm 3.5 $\\
020167131&	02:26:47.307&	$-$4:37:55.394&	1.2237&	3& D&	$10.08\pm0.19$&	$1.37\pm0.40$&	 $-50.0\pm 7.7 $&	$9.1\pm 1.8 $\\
020182331&	02:26:44.235&	$-$4:35:51.918&	1.2286&	3& D&	$10.72\pm0.11$&	$1.84\pm0.15$&	 $-95.5\pm 5.8 $&	$19.8\pm 1.4 $\\
020193070&	02:25:18.716&	$-$4:34:19.963&	1.0273&	3& D&	$10.15\pm0.20$&	$1.39^{+0.09}_{-0.20}$&	 $-44.9\pm 3.2 $&	$4.1\pm 0.3 $\\
020208482&	02:25:16.736&	$-$4:32:11.936&	1.0372&	2& D&	$10.17\pm0.16$&     $1.07\pm0.36$&	 $-55.3\pm 4.9 $&	$3.3\pm 0.4 $\\
020214655&	02:26:23.445&	$-$4:31:23.005&	1.0369&	2& D&	$10.02\pm0.16$&	$1.16\pm0.40$&	 $-59.6\pm 2.9 $&	$10.2\pm 0.6 $\\
020217890&	02:26:27.162&	$-$4:30:51.829&	1.5129&	3& D&	$9.99\pm0.19$&	$1.96^{+0.20}_{-0.12}$&	$\cdots$&	$\cdots$\\
020218856&	02:27:11.454&	$-$4:30:51.023&	1.3096&	9& D&	 $9.70\pm0.26$&	$1.09\pm0.40$&	 $-75.8\pm 6.3 $&	$7.8\pm 0.9 $\\
020239133&	02:26:42.981&	$-$4:28:30.821&	1.0199&	3& D&	 $9.89\pm0.15$&	$1.01\pm0.46$&	 $-82.6\pm 3.7 $&	$10.9\pm 0.4 $\\
020240675&	02:26:54.139&	$-$4:28:17.548&	1.3270&	3& D&	 $9.96\pm0.18$&      $1.16\pm0.51$&	 $-86.1\pm 11.5 $&	$11.5\pm 1.8 $\\
020255799&	02:26:45.858&	$-$4:26:15.767&	1.0352&	2& D&	 $9.87\pm0.16$&	$0.86\pm0.37$&	 $-74.8\pm 2.8 $&	$11.3\pm 0.5 $\\
020258016&	02:27:02.168&	$-$4:26:01.932&	1.3087&	2& D&	$10.61\pm0.12$&	$1.51\pm0.27$&	 $-48.4\pm 4.5 $&	$5.7\pm 0.6 $\\
020261328&	02:27:11.049&	$-$4:25:31.598&	1.5291&	2& D&	$10.01\pm0.20$&	$0.98^{+0.57}_{-0.39}$&	 $-180.0\pm 20.0$&	$10.0\pm 2.0$\\
020278667&	02:25:58.224&	$-$4:23:12.257&	1.0487&	9& D&	$10.28\pm0.17$&	$1.24\pm0.15$&	 $-56.1\pm 6.8 $&	$3.00\pm 0.4 $\\
020283083&	02:26:30.872&	$-$4:22:36.588&	1.2809&	2& D&	$10.05\pm0.21$&	$1.42\pm0.37$&	 $-76.2\pm 3.2 $&	$15.3\pm 0.9 $\\
020283830&	02:26:28.936&	$-$4:22:31.598&	1.3949&	3& D&	$10.37\pm0.17$&	$1.87\pm0.21$&	 $-49.5\pm 6.3 $&	$8.3\pm 1.2 $\\
020294045&	02:25:47.139&	$-$4:21:07.596&	1.0019&	3& D&	 $9.80\pm0.15$&	$0.89\pm0.45$&	 $-97.6\pm 2.1 $&	$16.5\pm 0.4 $\\
020306817&	02:25:50.316&	$-$4:19:22.933&	1.2225&	2& D&	 $9.76\pm0.21$&	$0.80^{+0.51}_{-0.45}$&	 $-144.3\pm 9.0 $&	$12.7\pm 1.0 $\\
020363717&	02:26:23.694&	$-$4:11:58.218&	1.3335&	3& D&	 $9.68\pm0.20$&	$1.41\pm0.31$&	 $-58.5\pm 4.3 $&	$30.7\pm 1.8 $\\
020370467&	02:26:14.699&	$-$4:11:05.428&	1.3330&	3& D&	$10.57\pm0.14$&	$1.70^{+0.06}_{-0.14}$&	 $-49.5\pm 3.2 $&	$10.9\pm 0.6 $\\
020386743&	02:27:13.988&	$-$4:08:59.564&	1.0465&	2& D&	 $9.88\pm0.21$&	$1.21\pm0.51$&	 $-67.6\pm 5.8 $&	$20.6\pm 2.5 $\\
020461235&	02:26:47.110&	$-$4:23:55.795&	1.0351&	4& D&	$10.36\pm0.14$&	$1.37\pm0.30$&	 $-68.2\pm 5.5 $&	$14.9\pm 1.5 $\\
020461893&	02:27:12.256&	$-$4:23:11.220&	1.0486&	2& D&	 $9.66\pm0.22$&	$1.05\pm0.39$&	 $-59.5\pm 2.2 $&	$22.3\pm 1.1 $\\
020465775&	02:26:59.371&	$-$4:18:59.994&	1.3580&	3& D&	$10.12\pm0.20$&	$1.47\pm0.41$&	 $-121.4\pm 23.3 $&	$7.7\pm 2.2 $\\
140083410&	13:57:50.629&	04:17:39.304&	0.9426&	3& W&	$10.07\pm0.18$&	$1.21\pm0.62$&	 $-70.0\pm 1.6 $&	$27.3\pm 1.0 $\\
140096645&	13:58:26.327&	04:19:47.784&	0.9623&	23& W&	$10.40\pm0.24$&	 $1.55\pm0.17$&	 $-195.6\pm 6.1 $&	$16.6\pm 0.6 $\\
140123568&	13:55:57.627&	04:24:20.095&	1.0016&	29& W&	 $9.73\pm0.40$&	$0.88\pm0.37$&	 $-65.0\pm 5.4 $&	$1.5\pm 0.2 $\\
140137235&	13:56:12.782&	04:26:31.798&	1.0444&	2& W&	$10.07\pm0.29$&	 $1.26\pm0.57$&	 $-59.4\pm 4.3 $&	$1.2\pm 0.1 $\\
140217425&	13:57:56.409&	04:38:37.086&	0.9758&	3& W&	$10.84\pm0.17$&	 $1.98\pm0.18$&	$-60.9\pm 1.4 $&	$25.4\pm 0.7 $\\
140258511&	14:00:19.667&	04:44:45.834&	1.2421&	3& W&	$10.80\pm0.48$&	$2.00\pm0.48$&	 $-74.2\pm 6.0 $&	$8.2\pm 0.8 $\\
140262766&	13:59:55.477&	04:45:30.150&	1.2813&	22& W&	 $9.84\pm0.43$&	 $0.79^{+0.68}_{-0.41}$&	 $-59.8\pm 4.5 $&	$5.4\pm 0.3 $\\
140545062&	13:59:35.586&	05:30:31.090&	1.0401&	2& W&	$10.60\pm0.18$&	$1.56\pm0.16$&	 $-28.5\pm 1.5 $&	$7.8\pm 0.5 $\\
220014252&	22:17:45.690&	00:28:39.468&	1.3097&	4& W&	$10.78\pm0.21$&	$1.94\pm0.42$&	 $-56.4\pm 2.6 $&	$37.5\pm 1.7 $\\
220015726&	22:15:42.455&	00:29:03.595&	1.3091&	2& W&	$10.77\pm0.27$&	 $2.05^{+0.11}_{-0.16}$&	 $-58.6\pm 6.8 $&	$9.5\pm 1.4 $\\
220071601&	22:18:01.569&	00:45:34.693&	1.3538&	2& W&	$10.81\pm0.57$&	 $2.47^{+0.08}_{-0.18}$&	 $-62.2\pm 6.3 $&	$25.6\pm 3.1 $\\
220148046&	22:14:37.803&	01:08:20.918&	2.2400$^1$&	2& W&	$11.22\pm0.16$&	 $3.20^{+0.02}_{-0.17}$&	 $\cdots $&	$\cdots $\\
220376206&	22:20:05.780&	00:08:21.682&	1.2440&	2& W&	$10.67\pm0.27$&	 $1.98\pm0.39$&	 $-74.9\pm 3.6 $&	$30.2\pm 1.3 $\\
220386469&	22:19:56.583&	00:03:03.161&	1.0240&	9& W&	$10.80\pm0.16$&	 $1.89\pm0.15$&	 $-71.8\pm 2.4 $&	$10.1\pm 0.3 $\\
220397579&	22:20:36.519&	00:01:46.682&	1.0367&	4& W&	$10.23\pm0.17$&	 $1.35\pm0.51$&	 $-114.0\pm 4.5 $&	$27.5\pm 1.6 $\\
220544103&	22:15:25.708&	00:06:39.535&	1.3970&	2& W&	$10.71\pm0.27$&	 $1.80\pm0.41$&	 $-55.4\pm 6.6 $&	$25.3\pm 3.0 $\\
220544394&	22:14:24.152&	00:06:46.897&	1.0077&	3& W&	$10.34\pm0.23$&	 $1.49\pm0.38$&	 $-49.2\pm 2.7 $&	$22.6\pm 1.9 $\\
220576226&	22:16:11.438&	00:16:30.436&	1.0217&	9& W&	$10.31\pm0.23$&	$1.62\pm0.44$&	 $-42.4\pm 5.7 $&	$11.1\pm 2.1 $\\
220578040&	22:17:04.103&	00:16:56.377&	1.0461&	2& W&	$10.72\pm0.16$&	 $1.57\pm0.21$&	 $-30.4\pm 2.7 $&	$14.8\pm 1.6 $\\
220584167&	22:15:23.038&	00:18:47.012&	1.4637&	3& W&	$11.21\pm0.25$&	 $2.23\pm0.25$&	 $-45.6\pm 6.2$&	$18.0\pm 2.0$\\
220596913&	22:14:29.184&	00:22:18.890&	1.2667&	2& W&	$10.68\pm0.29$&	 $2.00\pm0.30$&	 $-52.2\pm 4.8 $&	$18.5\pm 1.8 $\\
910154631&	02:27:28.736&	$-$4:38:00.254&	1.3327&	3& UD&	 $9.88\pm0.33$&	 $1.84\pm0.23$&	$-75.9\pm 6.8 $&	$6.9\pm 0.6 $\\
910159867&	02:26:43.634&	$-$4:37:09.098&	1.4823&	2& UD&	 $9.70\pm0.31$&	 $0.99^{+0.46}_{-0.39}$&	 $-45.8\pm 4.8 $&	$1.4\pm 0.2 $\\
910163602&	02:26:44.819&	$-$4:36:42.311&	1.3263&	3& UD&	$10.83\pm0.12$&	 $1.72\pm0.40$&	$-31.2\pm 3.2 $& $4.2\pm 0.4 $\\
910177382&	02:27:32.130&	$-$4:34:44.674&	1.2893&	4& UD&	$10.15\pm0.19$&	$1.39\pm0.16$&	$-85.0\pm 7.5 $&	$4.8\pm 0.4 $\\
910184233&	02:27:18.779&	$-$4:33:49.288&	1.4614&	2& UD&	 $9.55\pm0.35$&	  $0.90^{+0.45}_{-0.36}$&	 $-54.0\pm 4.2 $&	$2.1\pm 0.4 $\\
910186191&	02:26:17.891&	$-$4:33:34.027&	1.5399&	23& UD& $11.27\pm0.45$&   $2.22^{+0.60}_{-0.32}$&	$\cdots$&	$\cdots$\\
910187744&	02:27:17.235&	$-$4:33:20.052&	1.3191&	3& UD&	$10.07\pm0.20$&	 $1.44\pm0.18$& $-79.7\pm 7.2 $&	$5.6\pm 0.5 $\\
910191357&	02:25:26.617&	$-$4:32:51.025&	1.2863&	3& UD&	 $9.99\pm0.22$&	 $1.13\pm0.39$&	$-48.0\pm 4.5 $&	$1.3\pm 0.2 $\\
910193711&	02:25:46.278&	$-$4:32:34.066&	1.5523&	4& UD&	 $9.99\pm0.18$&	 $1.95\pm0.31$&	 $\cdots$&	$\cdots$\\
910195040&	02:27:32.118&	$-$4:32:24.490&	1.7712&	4& UD&	$10.58\pm0.20$&	 $1.86^{+0.58}_{-0.25}$&	 $\cdots$&	$\cdots$\\
910207502&	02:26:17.248&	$-$4:30:47.592&	1.4656&	4& UD&	$10.31\pm0.33$&	 $1.97^{+0.05}_{-0.18}$&	$-55.0\pm 5.1 $&	$4.0\pm 0.4 $\\
910224801&	02:25:37.690&	$-$4:28:32.606&	1.3201&	3& UD&	 $9.96\pm0.20$&	 $1.34\pm0.50$&$-56.3\pm 4.8 $&	$3.3\pm 0.3 $\\
910232719&	02:26:42.439&	$-$4:27:26.554&	1.7592&	3& UD&	$10.37\pm0.22$&	 $1.96\pm0.32$&	 $\cdots$&	$\cdots$\\
910238285&	02:27:16.371&	$-$4:26:44.682&	1.5405&	3& UD&	$10.92\pm0.13$&	 $1.89\pm0.18$&	 $\cdots$&	$\cdots$\\
910247797&	02:27:11.027&	$-$4:25:31.559&	1.5267&	3& UD&	 $9.77\pm0.35$&	 $1.30^{+0.53}_{-0.34}$&	 $\cdots$&	$\cdots$\\
910250031&	02:26:33.021&	$-$4:25:08.908&	1.6609&	2& UD&	$10.15\pm0.20$&	 $1.55\pm0.40$&	 $\cdots$&	$\cdots$\\
910254325&	02:27:29.241&	$-$4:24:40.136&	1.2818&	2& UD&	$10.22\pm0.27$&	$1.36\pm0.41$& $-38.8\pm 3.9 $&	$1.9\pm 0.2 $\\
910259245&	02:26:44.889&	$-$4:23:56.486&	1.5174&	3& UD&	$10.79\pm0.12$&	$1.87^{+0.06}_{-0.13}$&	 $\cdots$&	$\cdots$\\
910261247&	02:26:52.706&	$-$4:23:45.074&	1.4262&	4& UD&	$10.57\pm0.14$&	 $1.70\pm0.39$&	$-70.0\pm 6.5 $&	$4.0\pm 0.3 $\\
910262816&	02:27:13.437&	$-$4:23:29.756&	1.4807&	3& UD&	$11.26\pm0.15$&	 $1.50^{+0.04}_{-0.10}$&	$-65.4\pm 5.8 $&	$1.6\pm 0.2 $\\
910266034&	02:27:09.381&	$-$4:23:04.596&	1.5642&	4& UD&	$10.07\pm0.20$&	 $1.50\pm0.36$&	 $\cdots$&	$\cdots$\\
910274060&	02:25:48.031&	$-$4:21:58.842&	1.5680&	3& UD&	$10.09\pm0.21$&	 $1.32^{+0.55}_{-0.39}$&	 $\cdots$&	$\cdots$\\
910276733&	02:26:24.178&	$-$4:21:42.091&	1.3390&	4& UD&	$10.14\pm0.19$&	 $1.28\pm0.55$& $-66.6\pm 4.8 $&	$5.1\pm 0.5 $\\
910279515&	02:25:36.246&	$-$4:21:15.772&	1.3983&	3& UD&	$10.79\pm0.14$&	$1.90\pm0.07$&	 $\cdots$&	$\cdots$\\
910279755&	02:25:48.722&	$-$4:21:15.610&	1.3127&	4& UD&	 $9.96\pm0.24$&	 $1.34^{+0.45}_{-0.36}$&	 $-56.9\pm 5.7 $&	$5.8\pm 0.5 $\\
910286831&	02:26:17.403&	$-$4:20:15.061&	1.4696&	3& UD&	$10.28\pm0.27$&	 $1.82^{+0.05}_{-0.16}$&	 $-70.0\pm 6.0 $&	$3.8\pm 0.3 $\\
910296626&	02:25:50.696&	$-$4:19:00.556&	1.3570&	2& UD&	 $9.87\pm0.28$&	 $1.24\pm0.53$&	 $\cdots$&	$\cdots$\\
910300117&	02:27:27.438&	$-$4:18:33.124&	1.6791&	3& UD&	$10.27\pm0.31$&	 $1.22\pm0.52$&	 $\cdots$&	$\cdots$\\
910337228&	02:26:50.445&	$-$4:13:19.920&	1.3972&	3& UD&	 $9.96\pm0.20$&	 $1.34\pm0.40$&	$-107.4\pm 8.2 $&	$6.2\pm 0.5 $\\
910340496&	02:27:14.061&	$-$4:12:54.457&	1.3981&	3& UD&	$10.03\pm0.25$&	 $1.20\pm0.30$&	$-67.0\pm 5.6 $&	$3.2\pm 0.3 $\\
910360676&	02:27:10.203&	$-$4:10:25.082&	1.7230&	4& UD&	$10.56\pm0.14$&	 $2.37^{+0.18}_{-0.21}$&	 $\cdots$&	$\cdots$\\
910370574&	02:26:55.915&	$-$4:09:02.995&	1.6632&	2& UD&	$10.09\pm0.24$&	 $1.47^{+0.39}_{-0.43}$&	 $\cdots$&	$\cdots$\\
910371309&	02:27:12.473&	$-$4:08:59.158&	1.7926&	3& UD&	$10.44\pm0.30$&	 $2.13^{+0.06}_{-0.15}$&	 $\cdots$&	$\cdots$\\
910377628&	02:27:18.456&	$-$4:08:06.209&	1.4833&	2& UD&	$10.28\pm0.30$&	 $1.58^{+0.09}_{-0.15}$&	 $-110.7\pm 8.7 $&	$2.1\pm 0.2 $\\
\hline
\multicolumn{10}{l}{$^1$ For this galaxy, we indicate the redshift derived from \halpha\ in the SINFONI data (see text for details)}  
\end{longtable}
}


\begin{thebibliography}{}
\bibitem[Adelberger et al.(2004)]{2004ApJ...607..226A} Adelberger, K.~L., Steidel, C.~C., Shapley, A.~E., Hunt, M.~P., Erb, D.~K., Reddy, N.~A., \& Pettini, M.\ 2004, \apj, 607, 226 
\bibitem[Argence \& Lamareille(2009)]{2009A&A...495..759A} Argence, B., \& Lamareille, F.\ 2009, \aap, 495, 759
\bibitem[Bell et al.(2003)]{2003ApJS..149..289B} Bell, E.~F., McIntosh, D.~H., Katz, N., \& Weinberg, M.~D.\ 2003, \apjs, 149, 289 
\bibitem[Bernardi et al.(2010)]{2010MNRAS.404.2087B} Bernardi, M., Shankar, F., Hyde, J.~B., et al.\ 2010, \mnras, 404, 2087
\bibitem[Bonnet et al.(2004)]{2004SPIE.5490..130B} Bonnet, H., et al.\ 2004, \procspie, 5490, 130 
\bibitem[Bouch{\'e} et al.(2010)]{2010ApJ...718.1001B} Bouch{\'e}, N., et al.\ 2010, \apj, 718, 1001 
\bibitem[Bournaud et al.(2007)]{2007A&A...476.1179B} Bournaud, F., Jog, C.~J., \& Combes, F.\ 2007, \aap, 476, 1179
\bibitem[Bruzual \& Charlot(2003)]{2003MNRAS.344.1000B} Bruzual, G., \& Charlot, S.\ 2003, \mnras, 344, 1000 
\bibitem[Chabrier(2003)]{2003PASP..115..763C} Chabrier, G.\ 2003, \pasp, 115, 763 
\bibitem[Chary \& Elbaz(2001)]{2001ApJ...556..562C} Chary, R., \& Elbaz, D.\ 2001, \apj, 556, 562 
\bibitem[Conselice et al.(2008)]{2008MNRAS.386..909C} Conselice, C.~J., Rajgor, S., \& Myers, R.\ 2008, \mnras, 386, 909 
\bibitem[Conselice et al.(2009)]{2009MNRAS.394.1956C} Conselice, C.~J., Yang, C., \& Bluck, A.~F.~L.\ 2009, \mnras, 394, 1956 
\bibitem[Cucciati et al.(2011)]{2011arXiv1109.1005C} Cucciati, O., Tresse, L., Ilbert, O., et al.\ 2011, arXiv:1109.1005
\bibitem[de Ravel et al.(2009)]{2009A&A...498..379D} de Ravel, L., et al.\ 2009, \aap, 498, 379 
\bibitem[Dekel et al.(2009)]{2009Natur.457..451D} Dekel, A., et al.\ 2009, \nat, 457, 451 
\bibitem[]{} Eisenhauer, F., et al.\ 2003, Proc. SPIE, 4841, 1548 
\bibitem[Epinat et al.(2009)]{2009A&A...504..789E} Epinat, B., et al.\ 2009, \aap, 504, 789 
\bibitem[Epinat et al.(2011)]{} Epinat, B., et al.\ 2011, \aap, submitted
\bibitem[Erb et al.(2006)]{2006ApJ...647..128E} Erb, D.~K., Steidel, C.~C., Shapley, A.~E., Pettini, M., Reddy, N.~A., \& Adelberger, K.~L.\ 2006, \apj, 647, 128 
\bibitem[F{\"o}rster Schreiber et al.(2006)]{2006ApJ...645.1062F} F{\"o}rster Schreiber, N.~M., et al.\ 2006, \apj, 645, 1062 
\bibitem[F{\"o}rster Schreiber et al.(2009)]{2009ApJ...706.1364F} F{\"o}rster Schreiber, N.~M., et al.\ 2009, \apj, 706, 1364 
\bibitem[Franzetti et al.(2008)]{2008ASPC..394..642F} Franzetti, P., Scodeggio, M., Garilli, B., Fumana, M., \& Paioro, L.\ 2008, Astronomical Data Analysis Software and Systems XVII, 394, 642 
\bibitem[Garilli et al.(2008)]{2008A&A...486..683G} Garilli, B., et al.\ 2008, \aap, 486, 683 
\bibitem[Gavazzi et al.(2002)]{2002ApJ...576..135G} Gavazzi, G., Bonfanti, C., Sanvito, G., Boselli, A., \& Scodeggio, M.\ 2002, \apj, 576, 135 
\bibitem[Genel et al.(2008)]{2008ApJ...688..789G} Genel, S., et al.\ 2008, \apj, 688, 789 
\bibitem[Genel et al.(2010)]{2010ApJ...719..229G} Genel, S., Bouch{\'e}, N., Naab, T., Sternberg, A., \& Genzel, R.\ 2010, \apj, 719, 229 
\bibitem[Gnerucci et al.(2011)]{2011A&A...528A..88G} Gnerucci, A., et al.\ 2011, \aap, 528, A88 
\bibitem[Hopkins(2006)]{} Hopkins, A.~M..\ 2006, astro-ph/0611283
\bibitem[Ilbert et al.(2005)]{2005A&A...439..863I} Ilbert, O., Tresse, L., Zucca, E., et al.\ 2005, \aap, 439, 863
\bibitem[Iovino et al.(2005)]{2005A&A...442..423I} Iovino, A., et al.\ 2005, \aap, 442, 423 
\bibitem[Kennicutt(1998)]{1998ARA&A..36..189K} Kennicutt, R.~C., Jr.\ 1998, \araa, 36, 189 
\bibitem[Kere{\v s} et al.(2005)]{2005MNRAS.363....2K} Kere{\v s}, D., Katz, N., Weinberg, D.~H., \& Dav{\'e}, R.\ 2005, \mnras, 363, 2
\bibitem[Kere{\v s} et al.(2009)]{2009MNRAS.395..160K} Kere{\v s}, D., Katz, N., Fardal, M., Dav{\'e}, R., \& Weinberg, D.~H.\ 2009, \mnras, 395, 160
\bibitem[Lamareille et al.(2006)]{2006A&A...448..893L} Lamareille, F., Contini, T., Le Borgne, J.-F., Brinchmann, J., Charlot, S., \& Richard, J.\ 2006, \aap, 448, 893 
\bibitem[Lamareille et al.(2009)]{2009A&A...495...53L} Lamareille, F., et al.\ 2009, \aap, 495, 53 
\bibitem[Law et al.(2007)]{2007ApJ...669..929L} Law, D.~R., Steidel, C.~C., Erb, D.~K., Larkin, J.~E., Pettini, M., Shapley, A.~E., \& Wright, S.~A.\ 2007, \apj, 669, 929 
\bibitem[Law et al.(2009)]{2009ApJ...697.2057L} Law, D.~R., Steidel, C.~C., Erb, D.~K., Larkin, J.~E., Pettini, M., Shapley, A.~E., \& Wright, S.~A.\ 2009, \apj, 697, 2057 
\bibitem[Le F{\`e}vre et al.(2004)]{2004A&A...417..839L} Le F{\`e}vre, O., et al.\ 2004, \aap, 417, 839 
\bibitem[Le F{\`e}vre et al.(2005)]{2005A&A...439..845L} Le F{\`e}vre, O., et al.\ 2005, \aap, 439, 845 
\bibitem[Le Tiran et al.(2011)]{2011arXiv1104.4496L} Le Tiran, L., Lehnert, M.~D., Di Matteo, P., Nesvadba, N.~P.~H., \& van Driel, W.\ 2011, arXiv:1104.4496 
\bibitem[Lin et al.(2008)]{2008ApJ...681..232L} Lin, L., et al.\ 2008, \apj, 681, 232 
\bibitem[Liu et al.(2008)]{2008ApJ...678..758L} Liu, X., Shapley, A.~E., Coil, A.~L., Brinchmann, J., \& Ma, C.-P.\ 2008, \apj, 678, 758
\bibitem[Lonsdale et al.(2003)]{2003PASP..115..897L} Lonsdale, C.~J., et  al.\ 2003, \pasp, 115, 897
\bibitem[L{\'o}pez-Sanjuan et al.(2011)]{2011A&A...530A..20L} L{\'o}pez-Sanjuan, C., et al.\ 2011, \aap, 530, A20 
\bibitem[Maier et al.(2006)]{2006ApJ...639..858M} Maier, C., Lilly, S.~J., Carollo, C.~M., et al.\ 2006, \apj, 639, 858
\bibitem[Maiolino et al.(2008)]{2008A&A...488..463M} Maiolino, R., et al.\ 2008, \aap, 488, 463 
\bibitem[Mancini et al.(2011)]{2011arXiv1109.5952M} Mancini, C., Foerster Schreiber, N., Renzini, A., et al.\ 2011, arXiv:1109.5952
\bibitem[Mannucci et al.(2009)]{2009MNRAS.398.1915M} Mannucci, F., et al.\ 2009, \mnras, 398, 1915 
\bibitem[Meneux et al.(2008)]{2008A&A...478..299M} Meneux, B., Guzzo, L., Garilli, B., et al.\ 2008, \aap, 478, 299
\bibitem[Neichel et al.(2008)]{2008A&A...484..159N} Neichel, B., et al.\ 2008, \aap, 484, 159 
\bibitem[Ocvirk et al.(2008)]{2008MNRAS.390.1326O} Ocvirk, P., Pichon, C., \& Teyssier, R.\ 2008, \mnras, 390, 1326 
\bibitem[Osterbrock(1989)]{1989agna.book.....O} Osterbrock, D.~E.\ 1989, Research supported by the University of California, John Simon Guggenheim Memorial Foundation, University of Minnesota, et al.~Mill Valley, CA, University Science Books, 1989, 422 p. 
\bibitem[Pei(1992)]{1992ApJ...395..130P} Pei, Y.~C.\ 1992, \apj, 395, 130 
\bibitem[P{\'e}rez-Montero et al.(2007)]{2007MNRAS.381..125P} P{\'e}rez-Montero, E., H{\"a}gele, G.~F., Contini, T., \& D{\'{\i}}az, {\'A}.~I.\ 2007, \mnras, 381, 125 
\bibitem[P{\'e}rez-Montero \& Contini(2009)]{2009MNRAS.398..949P} P{\'e}rez-Montero, E., \& Contini, T.\ 2009, \mnras, 398, 949 
\bibitem[P{\'e}rez-Montero et al.(2009)]{2009A&A...495...73P} P{\'e}rez-Montero, E., et al.\ 2009, \aap, 495, 73 
\bibitem[Puech et al.(2007)]{2007A&A...466...83P} Puech, M., Hammer, F., Lehnert, M.~D., \& Flores, H.\ 2007, \aap, 466, 83 
\bibitem[Puech et al.(2008)]{2008A&A...484..173P} Puech, M., Flores, H., Hammer, F., et al.\ 2008, \aap, 484, 173
\bibitem[Puech et al.(2010)]{2010A&A...510A..68P} Puech, M., Hammer, F., Flores, H., Delgado-Serrano, R., Rodrigues, M., \& Yang, Y.\ 2010, \aap, 510, A68 
\bibitem[Queyrel et al.(2009)]{2009A&A...506..681Q} Queyrel, J., et al.\ 2009, \aap, 506, 681 
\bibitem[Queyrel et al.(2011)]{} Queyrel, J., et al.\ 2011, \aap, submitted
\bibitem[Ravikumar et al.(2007)]{2007A&A...465.1099R} Ravikumar, C.~D., et al.\ 2007, \aap, 465, 1099 
\bibitem[Salpeter(1955)]{1955ApJ...121..161S} Salpeter, E.~E.\ 1955, \apj, 121, 161 
\bibitem[Steidel et al.(2004)]{2004ApJ...604..534S} Steidel, C.~C., Shapley, A.~E., Pettini, M., Adelberger, K.~L., Erb, D.~K., Reddy, N.~A., \& Hunt, M.~P.\ 2004, \apj, 604, 534
\bibitem[Steidel et al.(2010)]{2010ApJ...717..289S} Steidel, C.~C., Erb, D.~K., Shapley, A.~E., Pettini, M., Reddy, N., Bogosavljevi{\'c}, M., Rudie, G.~C., \& Rakic, O.\ 2010, \apj, 717, 289
\bibitem[Vergani et al.(2008)]{2008A&A...487...89V} Vergani, D., et al.\ 2008, \aap, 487, 89 
\bibitem[Vergani et al.(2011)]{} Vergani, D., et al.\ 2011, \aap, submitted
\bibitem[Walcher et al.(2008)]{2008A&A...491..713W} Walcher, C.~J., et al.\ 2008, \aap, 491, 713 
\bibitem[Wisnioski et al.(2011)]{2011MNRAS.417.2601W} Wisnioski, E., Glazebrook, K., Blake, C., et al.\ 2011, \mnras, 417, 2601
\bibitem[Wright et al.(2007)]{2007ApJ...658...78W} Wright, S.~A., et al.\  2007, \apj, 658, 78 
\bibitem[Wright et al.(2009)]{2009ApJ...699..421W} Wright, S.~A., Larkin, J.~E., Law, D.~R., Steidel, C.~C., Shapley, A.~E., \& Erb, D.~K.\ 2009, \apj, 699, 421 
\bibitem[Yang et al.(2008)]{2008A&A...477..789Y} Yang, Y., et al.\ 2008, \aap, 477, 789 
\end{thebibliography}
\end{document}